\documentclass[prx,twocolumn,longbibliography,superscriptaddress,preprintnumbers]{revtex4-1}
\usepackage{amsmath}
\usepackage{amssymb}
\usepackage{booktabs} % For professional-quality table rules
\usepackage{array}    % For better column formatting
\usepackage{siunitx}  % For proper unit formatting (optional)
\usepackage{bm}
\usepackage{bbm}
\usepackage{braket}
\usepackage{amsthm}
\usepackage{diagbox}
\usepackage{setspace}
\usepackage{verbatim}
\usepackage{esint}
\usepackage{physics}
\usepackage{indentfirst}
\usepackage{graphicx}
\usepackage{float}
\usepackage{dsfont}
\usepackage{mathrsfs}
\usepackage{mathtools}
\usepackage{tikz-cd}
\usepackage{simplewick}
\usepackage{slashed}
\usepackage{mathtools}
\usepackage{hyperref}
\usepackage{slashed}
\usepackage{cancel}
\usepackage{xcolor}
\usepackage[normalem]{ulem}
\usepackage{youngtab}
\usepackage{ytableau}
\usepackage{enumerate}
\usepackage{leftindex}
\usepackage{svg}
\usepackage[T5]{fontenc}
\usepackage{cancel}
\usepackage{pst-node}

\usepackage{booktabs}
\usepackage{siunitx} % for 'S' column type
\usepackage{threeparttable} % for 'threeparttable' env.

\let \oldbm \bm
\renewcommand{\vec}[1]{\oldbm{#1}}

\def\bk{{\vec k}}
\def\bu{{\vec u}}
\def\bv{{\vec v}}

\def\ba{{\vec a}}

\def\bq{{\vec q}}
\def\bQ{{\vec Q}}

\def\bv{{\vec v}}
\def\bR{{\vec R}}
\def\bG{{\vec G}}

\def\bp{{\vec p}}

\def\bm{{\vec m}}

\def\br{{\vec r}}

\def\tr{\mathop{\mathrm{tr}}}
\def\Tr{\mathop{\mathrm{Tr}}}

\def\tr{\mathop{\mathrm{tr}}}

\def\O{\mathcal{O}}

\def\D{\mathcal{D}}
\def\Q{\mathcal{Q}}

\def\G{\mathcal{G}}

\def\H{\mathcal{H}}

\def\M{\mathcal{M}}

\def\diag{{\rm diag}}

\newcommand{\beq}{\begin{equation}}
\newcommand{\eeq}{\end{equation}}
\newcommand{\beqarray}{\begin{eqnarray}}
\newcommand{\eeqarray}{\end{eqnarray}}

\makeatletter
\newcommand\RedeclareMathOperator{%
  \@ifstar{\def\rmo@s{m}\rmo@redeclare}{\def\rmo@s{o}\rmo@redeclare}%
}
\newcommand\rmo@redeclare[2]{%
  \begingroup \escapechar\m@ne\xdef\@gtempa{{\string#1}}\endgroup
  \expandafter\@ifundefined\@gtempa
     {\@latex@error{\noexpand#1undefined}\@ehc}%
     \relax
  \expandafter\rmo@declmathop\rmo@s{#1}{#2}}
\newcommand\rmo@declmathop[3]{%
  \DeclareRobustCommand{#2}{\qopname\newmcodes@#1{#3}}%
}
\@onlypreamble\RedeclareMathOperator
\makeatother
\RedeclareMathOperator{\Im}{Im}
\RedeclareMathOperator{\Re}{Re}

\makeatletter
\let\oldabs\abs
\def\abs{\@ifstar{\oldabs}{\oldabs*}}
\let\oldnorm\norm
\def\norm{\@ifstar{\oldnorm}{\oldnorm*}}
\makeatother

\hypersetup{colorlinks=true,
            citecolor=blue,
            linkcolor=blue,
            urlcolor=blue,
            bookmarksnumbered=true}

\begin{document}

\title{Bootstrapping the Quantum Hall problem}
\author{Qiang Gao}
\affiliation{Department of Physics, Harvard University, Cambridge, Massachusetts 02138, USA}
\author{Ryan A. Lanzetta}
\affiliation{Department of Physics, University of Washington, Seattle, Washington 98195, USA}
\affiliation{Perimeter Institute for Theoretical Physics, Waterloo, Ontario N2L 2Y5, Canada}
\author{Patrick Ledwith}
\affiliation{Department of Physics, Harvard University, Cambridge, Massachusetts 02138, USA}
\author{Jie Wang}
\affiliation{Department of Physics, Temple University, Philadelphia, Pennsylvania, 19122, USA}
\author{Eslam Khalaf}
\affiliation{Department of Physics, Harvard University, Cambridge, Massachusetts 02138, USA}
\date{\today}
\begin{abstract}

The bootstrap method aims to solve problems by imposing constraints on the space of physical observables, which often follow from physical assumptions such as positivity and symmetry. Here, we employ a bootstrap approach to study interacting electrons in the lowest Landau level by minimizing the energy as a function of the static structure factor subject to a set of constraints, bypassing the need to construct the full many-body wavefunction. This approach rigorously lower bounds the ground state energy, making it complementary to conventional variational upper bounds. We show that the lower bound we obtain is relatively tight, within at most a few percent from the ground state energy computed with exact diagonalization (ED) at small system sizes, and generally gets tighter as we include more constraints. We also show that by combining the bootstrap lower bounds with variational Monte Carlo (VMC) calculations for various quantum Hall states, we can obtain two-sided bounds on the ground state energy that rigorously bounds the error to below a few percentage for large system sizes where ED is not accessible. Beyond the energetics, our results reproduce the correct power law dependence of the pair correlation function at short distances and the existence of a large entanglement gap in the two-particle entanglement spectra for the Laughlin states at $\nu = 1/3$. We further identify signatures of the composite Fermi liquid state close to half-filling. This shows that the bootstrap approach is capable, in principle, of describing non-trivial gapped topologically ordered, as well as gapless, phases.  We further study the evolution of the ground state energy and correlation functions as we interpolate between the lowest and first Landau levels at half-filling, and find indications of a phase transition from a composite Fermi liquid to a non-Abelian state. We also discuss more generally how the geometry of the allowed set of structure factors can be used to diagnose phase transitions. At the end, we will discuss possible extensions and current limitations of this approach and how they can be overcome in future studies. Our work establishes numerical bootstrap as a promising method to study many-body phases in topological bands, paving the way to its application in moir\'e platforms where the energetic competition between fractional quantum anomalous Hall, symmetry broken, and gapless states remains poorly understood.

\end{abstract}
\maketitle
\tableofcontents
\section{Introduction}
The importance of the quantum Hall problem in condensed matter physics cannot be overstated \cite{FQHRMP}. 
Not only did it provide one of the first experimental realizations of particle fractionalization and non-trivial statistics, it also led to the birth of the concept of topological order, which has since heavily influenced the study of phases of matter \cite{laughlin,PhysRevLett.53.722,PhysRevB.40.7387,Wen:1989iv}. Recently, the discoveries of both flat topological bands in moir\'e systems and, subsequently, fractional quantum Hall states at small \cite{XieTBGFCI} and zero magnetic fields \cite{MoTe2Xu2023, MoTe2MakShan2023, Xu2023Transport, LiFCIPRX, Pentalayer} have revived the study of quantum Hall phases. While the universal aspects of these zero-field fractional quantum Hall states are expected to be broadly similar to their finite-field counterparts, their detailed energetics and competition with other phases such as gapless metallic or symmetry-broken phases are quite different, requiring the application of unbiased numerics to capture such competition.

Most numerical studies of strongly correlated phases focus on spin lattice models or models of hopping fermions on a lattice e.g. the Hubbard model, where locality plays an important role. In contrast, the quantum Hall problem, and more generally the problem of interactions in topological flat bands, inherently lacks such a local lattice description. Furthermore, quantum Hall systems realize many exotic, strongly correlated phases that cannot be captured by Hartree-Fock or related mean-field methods.
As a result, numerical studies of the quantum Hall problem have been mostly restricted to exact diagonalization (ED) and density matrix renormalization group (DMRG) \cite{Zaletel_DMRG}. The former is limited by exponential scaling in system size, whereas the latter is restricted to cylinder geometries, with exponential scaling in the cylinder circumference, and is best suited to capture gapped phases with relatively low entanglement. This motivates the search for alternative numerical approaches that can overcome these limitations, complement existing methods, and exploit the rich structure of the problem. \par 

One class of such alternative approaches follows the \emph{bootstrap philosophy}. Broadly speaking, the bootstrap way of thinking aims to constrain the space of physical observables by exploiting symmetries and other physical assumptions, such as unitarity. Bootstrap ideas in this spirit were originally invoked to study scattering amplitudes \cite{chew_S_matrix}, and were central to the development of conformal field theory (CFT) \cite{kad_operator_alg,Ferrara:1972kab,Polyakov:1974gs, BPZ,Shenker_min_models}.
The recent introduction of numerical optimization techniques in conformal bootstrap \cite{Rattazzi:2008pe}---especially semidefinite programming (SDP) \cite{Poland:2011ey,wolkowicz2012handbook}---has revitalized the utility of the bootstrap philosophy, in particular providing a tractable and quantitative way to leverage positivity conditions~\cite{Note1}. Of the many interesting CFT results this new ingredient has produced, the precise determination of the critical exponents of the 3$d$ Ising CFT stands out as one of the clearest demonstrations of the potential power of bootstrap \cite{solving_3d_Ising,precision_islands}. The success of the modern numerical conformal bootstrap has motivated interest in exploring broader classes of problems benefiting from a bootstrap approach. Positivity conditions are abundant in quantum and statistical physics problems, and many recent works have shown that such conditions fit nicely within the modern numerical bootstrap framework \cite{HanBootstrappingMatrixQM, Berenstein:2021dyf, PhysRevD.108.125002, berenstein2022bootstrapping,han2020quantum,PhysRevD.107.L051501, cho2022bootstrapping,PhysRevD.106.116008, nancarrow2023bootstrapping}. We also note here several earlier and parallel works with a similar flavor to the now-developing quantum mechanics bootstrap program, including the pioneering work of Maziotti \cite{Mazziotti1998, Maziotti2002, Maziotti2004, Maziotti2005, Mazziotti2012, Mazziotti2020, Mazziotti2023} in quantum chemistry, as well as more general reduced density matrix (RDM) and semidefinite relaxation approaches \cite{massaccesi2021variational, SemidefiniteRelaxation, haim2020variational,baumgratz2012lower} applied to quantum many-body problems.

In this work, we employ a bootstrap approach to study density-density interactions in the lowest Landau level (LLL). 
We show that the energetics and correlation functions of known states can be reproduced to reasonable accuracy using significantly less computational resources relative to ED, due to the polynomial scaling of the method in system size. Our approach is based on imposing mathematical consistency conditions on the static structure factor $S_\bq := \langle \rho_\bq \rho_{-\bq} \rangle$, where $\rho_\bq$ is the guiding center density operator. For density-density interactions, the energy is completely determined by $S_\bq$. Importantly, since the $S_\bq$ of the exact ground state is necessarily part of the search space, minimizing the energy subject to the constraints yields a lower bound on the energy. The constraints we use include a self-duality first derived by Haldane \cite{HaldaneDuality}, combined with a set of positivity conditions. The latter are a subset of the $N$-representability conditions --- a set of necessary and sufficient conditions that ensure a given reduced density matrix (RDM) is derivable from a valid $N$-fermion density matrix \cite{Coleman1963, Erdahl1978, Mazziotti2012, Mazziotti2012PRL, Mazziotti2023} --- which we apply to the Landau level problem, exploiting its additional structure.

Let us emphasize several advantages of this method that makes it very promising for a wide range of many-body problems. First, while the only rigorous guarantee we have is a lower bound on the ground state energy, our results indicate that we can reproduce several features of the correlations functions in the ground state to very good accuracy; even for strongly correlated ground states with no mean-field description such as topologically-ordered gapped states and strongly-correlated gapless composite Fermi liquids. The rigorous lower bound also makes the method complementary to variational methods like Hartree Fock and DMRG which produce upper bounds. Second, while this method is a general method for any quantum many-body system, it is particularly powerful at exploiting problem-specific structures, and can incorporate symmetries and non-trivial analytical information about the problem through the selection of constraints. Third, by making connections to the bootstrap program, the approach can be extended to extract information about other quantities such as gaps, excitation spectra, etc. Finally, the method can be implemented fully in momentum space or any other basis, making it insensitive to the absence of a local lattice description in topological bands. 

\section{A brief review of Reduced Density Matrix optimization method}
Our work aims to employ a bootstrap approach to problems of fermions interacting with few body-interactions projected to a band in the continuum. While our problem has some special structures resulting from the band projection, it shares the same basic structure as the problem of electrons interacting via two-body interactions in vaccuum---an extensively-studied problem in quantum chemistry. Thus, our approach will heavily use the positivity constraints developed in quantum chemistry in the context of RDM methods (also called wavefunction-free methods) \cite{Mazziotti1998, Maziotti2002, Maziotti2004, Maziotti2005, Mazziotti2012, Mazziotti2020, Mazziotti2023}. In the following, we give a brief review of the general theory for such RDM positivity constraints.

We consider a system with $N$ fermions interacting with at most $p$-body interactions with $p \ll N$. In this work, we will focus on two-body interactions ($p=2$) which is the most general form of interaction obtained when projecting Coulomb or screened Coulomb interaction onto a set of bands, ignoring band mixing effects. Generalization to $p>2$ is straightforward. For $p=2$, the Hamiltonian of interest can always be written as:
\begin{equation}
    \H = \sum_{ij;kl}H^{ij}_{kl}c^\dagger_i c^\dagger_j c_l c_k,
\end{equation}
where the rank-$(2,2)$ tensor $H^{ij}_{kl}$ is anti-symmetric in the indices $i,j$ and $k,l$, respectively. In the following, we will use caligraphic letters to denote second quantized operators and the corresponding non-caligraphic letters to denote their first quantized version. Since we are acting on the Hilbert space of fixed particle number, we can absorb any single particle terms  into the two-body interaction through:
\begin{equation}\label{single_particle_terms}
    \sum_{ik}H_{ik}c^\dagger_i c_k = \sum_{ij;kl}\frac{1}{N-1}\mathcal{A}[H_{ik}\delta_{jl}]c^\dagger_i c^\dagger_j c_l c_k,
\end{equation}
where the operation $\mathcal{A}$ anti-symmetrizes the indices $i,j$ and $k,l$, respectively, and $N-1$ is the particle number of the system acted on by one annihilation operator. 

\subsection{Reformulating the many-body problem as an optimization problem}

The quantum many-body bootstrap relies on the observation that the wavefunction, which is a complicated object living in an exponentially large Hilbert space, is not needed to compute all the observables of a system. For instance, for a $2$-body interaction, only the $2$-particle reduced density matrix is needed to compute the energy which is given by
\begin{equation}
    E = \Tr[H \leftindex^2D]
\end{equation}
where $\leftindex^p D$ is the $p$-particle reduced density matrix ($p$-RDM) obtained from the full $N$-body density matrix $\leftindex^N D$ by tracing out $N-p$ particles. For $p=2$, this gives the 2-RDM
\begin{equation}
\leftindex^2 {\D}^{i_1i_2}_{j_1j_2} \equiv \langle c_{i_1}^\dagger c_{i_2}^\dagger c_{j_2} c_{j_1} \rangle 
\end{equation}
This means that we can find the ground state energy and all two-particle correlation functions by minimizing the functional $E[\leftindex^2D]:= \Tr [ H \leftindex^2D]$ over the space of 2-RDMs that can be obtained from an $N$-body fermionic density matrix. 

To phrase this problem more concretely, let us define the contraction or partial trace operator $L^p_N$ which maps an $N$-particle  density matrix to a $p$-particle density matrix by tracing over $N - p$ indices. We can then find the ground state energy and correlation functions by solving the optimization problem
\begin{gather}
    E_{\rm GS} = \min_{\leftindex^2D\in  P^2_N} \Tr[H\leftindex^2D] \\
    P^2_N = \{\leftindex^2D: \leftindex^2D = L_N^2 (\leftindex^ND)\}
\end{gather}
where $\leftindex^ND$ is a valid $N$-particle density matrix with the standard form $\leftindex^ND = \sum_i p_i |\psi_i \rangle \langle \psi_i|$ with $p_i \geq 0$, $\sum_i p_i = 1$ and $\hat N|\psi_i \rangle = N |\psi_i \rangle$. The energy functional to be optimized is a simple linear functional of $\leftindex^2D$ which lives in a space whose dimension scales polynomially with the single particle Hilbert space dimension. However, characterizing the set $P^2_N$, known as the $N$-representability problem \cite{Coleman1963}, turns out to be a hard problem; determining whether a 2-RDM belongs to $P^2_N$ was proven to be QMA-hard \cite{RN436}. This is not surprising since we know the many-body problem is generally hard, so in any formulation the difficulty has to appear somewhere. 

We note here the crucial distinction between the 1-RDM, which is used as the basis for the Hartree-Fock method, and the $p$-RDM for $p>1$. A matrix $D^i_j$ which is positive, hermitian, and properly normalized can always be interpreted as the 1-RDM for a (pure or mixed) density matrix. However, positivity, hermiticity and antisymmetry are necessary but not sufficient conditions for a $p$-RDM, $p>1$, to be obtainable from a valid $N$-body fermionic state \cite{Coleman1963}. The RDM method aims to approximate the set $P^2_N$ by imposing a partial set of necessary constraints that can be efficiently checked and yield a superset $\tilde P^2_N \supseteq  P^2_N$. Since any $\leftindex^2 D \in P^2_N$ satisfies all the constraints, the ground state 2-RDM is always contained in the search space. Thus, the energy $\tilde E_{\rm GS}$ obtained from the minimization procedure over the space $\tilde P^2_N$ is always a {\it lower bound} on the actual ground state energy, $\tilde E_{\rm GS} \leq E_{\rm GS}$. The set $\tilde P^2_N$ can be constructed as a convex set so that convex combinations of constraints generates new constraints. This allows the many-body bootstrap problem to be formulated as a convex optimization problem.

\subsection{Hierarchy of constraints}\label{2_p_constraints}
\begin{figure}
    \centering
    \includegraphics[width=\linewidth]{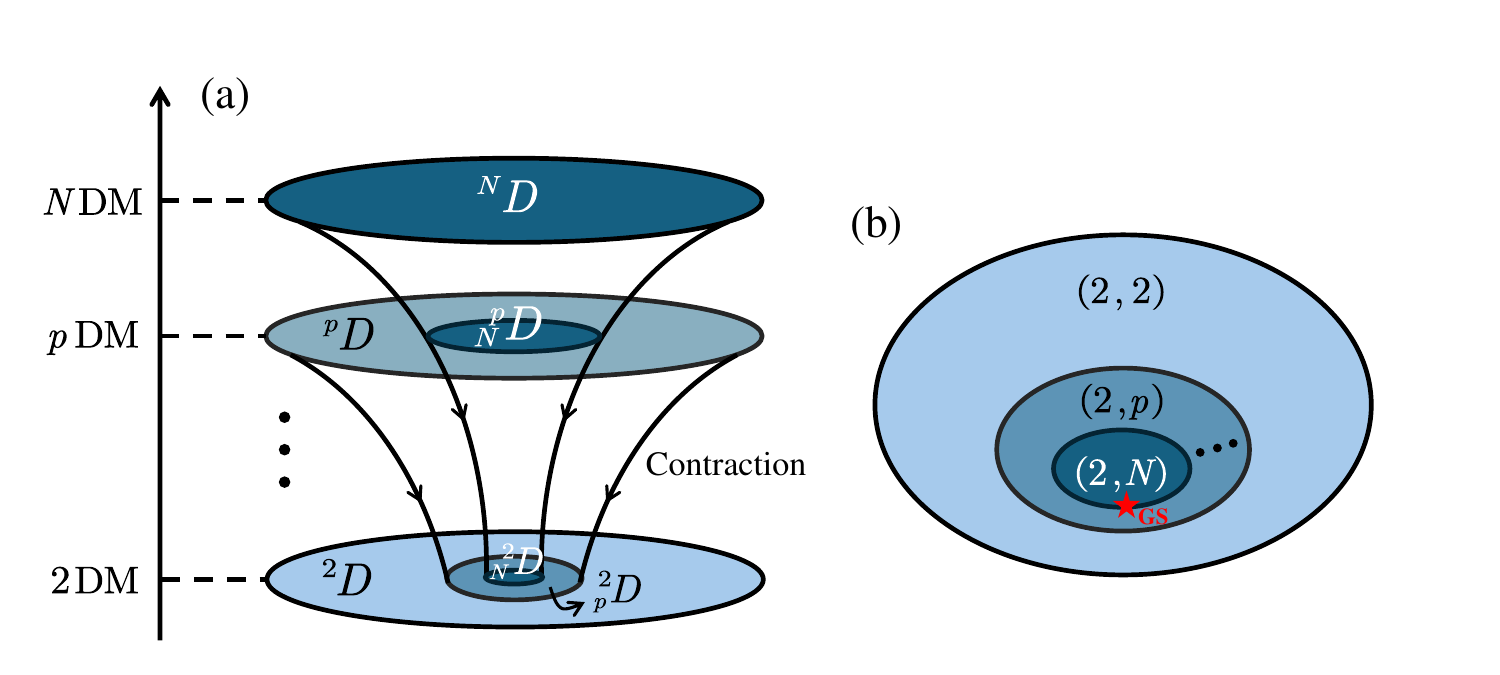}
    \caption{Hierarchy of density matrices (DMs) at different particle levels (a) and corresponding positivity constraints for the $N$-representable 2RDM (b). The red star in panel b denotes the true ground state.}
    \label{fig:Hierarchy}
\end{figure}
In conformal bootstrap, there is an infinite number of constraints whereas in quantum many-body bootstrap the number of constraint is finite but exponentially large for finite system size. In both cases, it is desirable to have a {\it hierarchy} of constraints that allows us to include the most important constraints first and then add more and more constraints until some convergence is achieved. In local lattice models, the hierarchy of constraints is usually based on locality \cite{haim2020variational} starting with constraints built from local operators and gradually incorporating operators acting on larger regions. In our problem, the projected interaction is not strictly local and there is no local lattice basis due to band topology. On the other hand, our interaction is a 2-body interaction and we expect the most important correlations to be few-body correlations. Thus, we will employ the hierarchy of constraints adapted from quantum chemistry as we will explain below.

 A hierarchy of constraints characterizing the set of $N$-representable 2-RDMs, $P^2_N$, can be constructed using the simple observation that any $N$-representable 2-RDM can be obtained by a process of contraction or partial trace of an $N$-representable $p$-RDM for $p>2$. Since the set of $N$-representable $p$-RDMs is a subset of the set of all $p$-DMs, $N$-representability implies $p$-representability, $P^2_N \subseteq P^2_p$~\cite{Note2}. This allows a gradation or hierarchy of constraints which approximates the full set of $N$-represantability constraints with partial sets of constraints for $p < N$, schematically illustrated in Fig.~\ref{fig:Hierarchy}a, which can be written as $P^2_N \subseteq P^2_{N-1} \subseteq P^2_{N-2} \dots P^2_{3} \subseteq P^2_{2}$.

The main disadvantage of the above procedure is that it requires constructing the higher $p$-RDMs. In Ref.~\cite{Mazziotti2012PRL}, Mazziotti showed that we can sidestep the need to construct higher RDMs through an explicit construction of the $N$-representability constraints that only depends on the 2-RDM. Maziotti's construction generalizes earlier constructions \cite{Garrod1964Reduction, Erdahl1978Representability} of positivity constraints and follows from the simple inequality
\begin{equation}
    \sum_i \langle \O_i \O_i^\dagger \rangle \geq 0
    \label{OOdIneq}
\end{equation}
for any choice of $\O_i$. For a general set of $p$-fermion operators $\{\O_i\}$, the summation above is expressible in terms of the $p$-RDM. However, there are special choices of $\{\O_i\}$ where the $r$-particle parts, $r > 2$, cancel allowing us to express (\ref{OOdIneq}) as a constraint on the 2-RDM. The constraints constructed from sets $\{\O_i \}$ where $\O_i$ is a polynomial in fermionic creation/annihilation operator of order at most $p$ are called $(2,p)$ constraints. Clearly from this definition, the set of $(2,p)$ constraints contains the set of $(2,p')$ constraints for $p' < p$~\cite{Note3}. This gives rise to a hierarchy of constraints which gets stronger as $p$ is increased as shown in Fig.~\ref{fig:Hierarchy}b. Maziotti \cite{Mazziotti2012PRL} has shown that \emph{all} $N$-representability constraints take this form. Thus, imposing all $(2,N)$ constraints is necessary and sufficient to determine whether a 2-RDM is $N$-representable.

The simplest constraints are constructed by taking $\O$ to be a 2-body operator. In that case, we do not need to sum over multiple terms in (\ref{OOdIneq}) and we get the positivity constraints (usually called the DQG conditions~\cite{Garrod1964Reduction}):
\begin{equation}
    \langle \mathcal{O}_D \mathcal{O}_D^\dagger \rangle \geq 0,\quad \langle \mathcal{O}_Q \mathcal{O}_Q^\dagger \rangle \geq 0, \quad \langle \mathcal{O}_G \mathcal{O}_G^\dagger \rangle \geq 0
\end{equation}
with $\mathcal{O}_D = \sum_{\alpha,\beta}A^D_{\alpha\beta}c^\dagger_\alpha c^\dagger_\beta$, $\mathcal{O}_Q = \sum_{\alpha,\beta}A^Q_{\alpha\beta}c_\alpha c_\beta$, and $\mathcal{O}_G = \sum_{\alpha,\beta}A^G_{\alpha\beta}c^\dagger_\alpha c_\beta$, for arbitrary $A^\text{D,Q,G}$ under proper anti-symmetrizations. All these constraints has the form $ [A^l]^\dagger \M^l(\leftindex^2D) A^l \geq 0$ for all $A^l$ where $l = D, Q, G$ (here we used a vector notation where the two lower indices of $A^l_{\alpha \beta}$ are combined together and $A^l$ is treated as a vector in this combined index). This translates to the condition that the matrices $\M^l(\leftindex^2D)$ which depend only on the 2-RDM is positive semidefinite (PSD), usually denoted by $\M^l(\leftindex^2D) \succeq 0$. Note that the first constraint just reflects the fact the the 2-RDM is a PSD matrix whereas the other two generally give different non-trivial constraints on the 2-RDM.

For $p>2$, a more careful choice of the $\mathcal O_i$ is required to yield constraints involving only the 2-RDM. To illustrate this, consider the case of $p = 3$. The simplest positivity constraint at this level, called T1~\cite{Erdahl1978Representability}, is defined through $\sum_{i=1}^2 \langle \O_{T1,i} \O^\dagger_{T1,i} \rangle \geq 0$ where $\O_{T1,1} = \sum_{\alpha\beta\gamma}A_{\alpha\beta\gamma}c^\dagger_\alpha c^\dagger_\beta c^\dagger_\gamma $ and $\O_{T1,2} = \O_{T1,1}^\dagger = \sum_{\alpha\beta\gamma}A^*_{\alpha\beta\gamma}c_\gamma c_\beta c_\alpha$. This constraint can be phrased as $\M^{T1} \succeq 0$ where  $\M^{T1}$ is defined as $[\M^{T1}]^{\alpha\beta\gamma}_{\alpha'\beta'\gamma'} := \langle \{c^\dagger_{\alpha} c^\dagger_\beta c^\dagger_\gamma, c_{\gamma'} c_{\beta'} c_{\alpha'}\}\rangle$. Note that when commuting $ccc$ past $c^\dagger c^\dagger c^\dagger$, the three-particle term cancels due to the fermionic anti-commutation relation. As a result, $\M^{T1}$ only contains expectations values of at most four fermions, leading to a PSD constraint on the 2-RDM of the form $\M^{T1}(\leftindex^2D) \succeq 0$. Another constraint we can construct at this level, called the (generalized) T2 constraint~\cite{Zhao2004RDM,Mazziotti2012}, is obtained by imposing $\sum_{i=1}^{2}\langle \O_{T2,i}\O_{T2,i}^\dagger\rangle \geq 0$
with $\O_{T2,1}\equiv\sum_{\alpha\beta\gamma}A_{\alpha\beta\gamma}c^\dagger_\alpha c^\dagger_\beta c_\gamma + \sum_\alpha B_\alpha c^\dagger_\alpha $ and $\O_{T2,2}\equiv\sum_{\alpha\beta\gamma}A^*_{\alpha\beta\gamma}c^\dagger_\gamma c_\beta c_\alpha   +\sum_\alpha D^*_\alpha c_\alpha$. Similar to the case of T1, it is easy to see that the expectation values containing more than four fermion operators cancel and we obtain a PSD constraint of the form $\M^{T2}(\leftindex^2D) \succeq 0$. Other constraints for $p = 3$ and $p > 3$ can be constructed similarly \cite{Mazziotti2012, Mazziotti2012PRL}, but they require imposing a tensor decomposition structure on the variables $A$, e.g. $A_{\alpha \beta \gamma} = B_{\alpha \beta} C_\gamma$. We will come back to this in the later discussions. For the rest of the manuscript, we will focus on the $(2,2)$, T1 and generalized T2 constraints.

\section{Problem setup}\label{problem_setup}
In this section, we will discuss the main bootstrap setup applied to the quantum Hall problem. We will find that the RDM positivity constraints take a particularly simple form in the quantum Hall problem due to continuous magnetic translation symmetry. In particular, the 2-RDM can be expressed in terms of the much simpler static structure factor.

\subsection{Interacting quantum Hall system}
We begin with a brief review of the basic setup of the quantum Hall problem. We consider the LLL on a torus with $N$ electrons and $N_\Phi$ flux quanta corresponding to the filling $\nu = N/N_\Phi$. For simplicity, we will set the magnetic length $l_B$ to $1$. A LLL projected density-density interaction $V_\bq$ is described by the Hamiltonian
\begin{equation}
    \H = \frac{1}{2} \int \frac{d^2 \bq}{2\pi} V_\bq \tilde \rho_{\bq} \tilde \rho_{-\bq}, \qquad \tilde \rho_\bq = \sum_{\alpha,\beta} \tilde \Lambda_\bq^{\alpha \beta} c_\alpha^\dagger c_\beta.
    \label{Hamiltonian}
\end{equation}
Here the sum over $\alpha$ and $\beta$ runs over any complete orthonormal basis of the LLL and $\tilde \Lambda_\bq^{\alpha \beta} = \langle \alpha|e^{-i \bq \cdot \br}|\beta \rangle$ is the form factor matrix.  We can write the LLL projected density operator as $\tilde \rho_\bq = e^{-\frac{\bq^2}{4}} \rho_\bq$, where $\rho_\bq = \sum_{\alpha,\beta} \Lambda_\bq^{\alpha \beta} c_\alpha^\dagger c_\beta$ with $\Lambda_\bq^{\alpha \beta} = \langle \alpha|e^{-i \bq \cdot \bR}|\beta \rangle$ and $\bR$ is the guiding center coordinate satisfying $[R_a, R_b] = -i \epsilon_{ab}$.
Generalization to the $n$-th LL can be straightforwardly implemented via the replacement $V_\bq \mapsto V_\bq [L_n(\bq^2/2)]^2$ where $L_n(x)$ denotes the $n$-th Laguerre polynomial. We note that on a system with $N_\Phi$ flux quanta placed on an $N_x \times N_y$ grid (with $N_\Phi = N_x N_y$), the momenta $\bq$ will be discretized in units of $\Delta_{x,y} = \frac{2\pi}{N_{x,y}}$ and the operators $\rho_\bq$ satisfy $\rho_{\bq + N_\Phi \bp} = \pm \rho_\bq$ for any $\bp$ on the grid \cite{HaldaneDuality}. This means that there are at most $N_\Phi^2$ distinct momenta. For the LLL, the dimension of the linear vector space spanned by $\rho_\bq$ for different $\bq$'s turns out to be exactly equal to $N_\Phi^2$. The momentum integral in Eq.~\eqref{Hamiltonian} and all subsequent expressions should be interpreted as $\int \frac{d^2 \bq}{2\pi} = \frac{1}{N_\Phi} \sum_\bq$ where the sum goes over these $N_\Phi^2$ distinct momenta~\cite{Note4}. For example, we can take $\bq = (n_x \Delta_x, n_y \Delta_y)$ where $n_x, n_y = 0,\dots,N_\Phi - 1$. 

The energy of any state for a density-density interaction is completely specified by the static structure factor, defined as $S_\bq = \langle \rho_\bq \rho_{-\bq} \rangle$:
\begin{equation}
    E[S_\bq] = \frac{1}{2} \int \frac{d^2 \bq}{2\pi} V_\bq e^{-\frac{\bq^2}{2}} S_\bq
    \label{ESq}
\end{equation}
where $S_\bq \geq 0$ and satisfies $S_{\bq = 0} = N^2$. It is also useful to define the structure factor per flux quantum \cite{HaldaneDuality} whose $\bq = 0$ component is subtracted off, which we will denote by $s_\bq := \frac{1}{N_\Phi} (S_\bq - N^2 \delta_{\bq,0})$. Many of our expressions are simpler when expressed in terms of $S_\bq$, but we will present the final numerical results always in terms of $s_\bq$, which does not scale with system size, to facilitate comparison with previous work. 

The structure factor encodes the full information about all two-body correlation functions which means it contains the same information as the 2-RDM. This may seem surprising since the 2-RDM is an antisymmetric matrix with linear dimension $\frac{N_\Phi (N_\Phi - 1)}{2}$ parametrized by $\left(\frac{N_\Phi (N_\Phi - 1)}{2}\right)^2 \sim N_\Phi^4$ variables. On the other hand, there are only $N_\Phi^2$ possible values for $\bq$. The resolution for this discrepancy arises from continuous magnetic translation symmetry (CMTS) which allows us to diagonalize the 2-RDM of any state~\cite{Note5} using a unitary transformation that does not depend on the state, as we will discuss later. This reduces the number of independent variables in the 2-RDM to the number of its eigenvalues, $\frac{N_\Phi (N_\Phi - 1)}{2}$. On the other hand, not all $N_\Phi^2$ values of $S_\bq$ are independent. In fact, a duality derived by Haldane \cite{HaldaneDuality}, which we will explain in detail, relates $S_\bq$ to its own Fourier transform reducing the number of independent values of $S_\bq$ to $\frac{N_\Phi (N_\Phi - 1)}{2}$ matching the count for the 2-RDM. Thus, we have two equivalent representations for the two-body correlation functions in the LLL:
\begin{equation}\label{s_qto_2RDM_equivalence}
    S_\bq + \text{Haldane duality} \Leftrightarrow \text{2RDM} + \text{CMTS}
\end{equation}
With this identification, we can write the $(2,p)$ positivity constraints, usually expressed in terms of the 2-RDM, in terms of $S_\bq$. It should be noted that using the $S_\bq$ representation makes the symmetries more manifest. For example, inversion symmetry has the simple implementation $S_\bq = S_{-\bq}$. 
Furthermore, in the thermodynamic limit, the full rotational symmetry implies $S_\bq = S_{|\bq|}$. Since we are always working on a torus, we will only have some discrete rotation symmetry depending on the chosen lattice. Here, we choose a square lattice with a $C_4$ such that $S_\bq = S_{R_4\bq}$ where $R_4$ is the $\pi/2$ rotation matrix. Note that as we increase system size, we should expect $S_\bq \approx S_{|\bq|}$ for sufficiently small $\bq$ such that the effect of boundary conditions can be neglected. 

We emphasize that because the search space of $S_\bq$'s (or more generally RDMs) is convex, the minimum energy always lies at its boundary. Combined with the fact that we are searching a larger region than the physical space, this means that the optimal $S_\bq$ will almost always not correspond to any physical state. However, if we have imposed enough constraints such that the search space is a good approximation for the physical region at least close to the minimum of the Hamiltonian, we expect the correlation functions, e.g. structure factor, we obtain to be a good approximation for the correlation functions of the true ground state.

\subsection{Haldane duality}
An important identity satisfied by the structure factor $S_\bq$ is a duality derived by Haldane \cite{HaldaneDuality}:
\begin{equation}\label{Haldane_duality}
    S_\bp - N = \pm \int \frac{d^2\bq}{2\pi} e^{i\bq\wedge\bp}(S_\bq-N).
\end{equation}
where the $\pm$ signs correspond to bosons and fermions, respectively, and the wedge product is defined as $\bq \wedge \bp = q_x p_y - q_y p_x$. We will be focusing exclusively on the fermionic case in this work. 

Haldane's proof of Eq.~\eqref{Haldane_duality} uses the properties of magnetic translation symmetry and its relation to the guiding center coordinates, which are features unique to the quantum Hall problem making it unclear whether this duality is generalizable beyond the LLL case. Here, we will provide an alternative proof that shows that the duality can be generalized to a much wider class of bands. The main assumption we will need is that the relation between the operator $\rho_\bq = e^{\frac{1}{4}\bq^2} \tilde \rho_\bq$ (which can be defined for any band) and the fermion bilinear $c_\alpha^\dagger c_\beta$ can be inverted:
\begin{gather}
    \rho_\bq = \sum_{\alpha,\beta} \Lambda^{\alpha \beta}_\bq c^\dagger_\alpha c_\beta \quad \Leftrightarrow\quad c^\dagger_\alpha c_\beta = \int \frac{d^2\bq}{2\pi} \Lambda^\bq_{\alpha \beta} \rho_\bq, \nonumber \\
    \sum_{\alpha,\beta} \Lambda^{\alpha \beta}_\bq \Lambda^{\bq'}_{\alpha \beta} = N_\Phi \delta_{\bq,\bq'}, \quad \int \frac{d^2\bq}{2\pi} \Lambda^\bq_{\alpha \beta} \Lambda_\bq^{\gamma \delta} = \delta_{\alpha,\gamma} \delta_{\beta,\delta}
    \label{invertability_form_factor}
\end{gather}
For an arbitrary band, lattice translation symmetry implies momentum conservation modulo reciprocal lattice translations which implies $\langle \rho_\bq \rho_\bp \rangle = \delta_{\bq+\bp,\bG} S_{\bq,\bG}$ where $\bG$ is a reciprocal lattice vector (for our grid, this corresponds to the vectors $\bG = 2\pi (n,m)$ for $n=0,\dots,N_y-1$ and $m=0,\dots,N_x-1$). In the LLL, CMTS implies full momentum conservation such that $S_{\bq,\bG} = \delta_{\bG,0} S_{\bq}$. Below, we will show that there is a more general version of the Haldane duality valid for any band satisfying the form factor invertibility relation (\ref{invertability_form_factor}) which reduces to the Haldane duality when using CMTS and the precise form of the matrices $\Lambda$ for the LLL. To derive this relation, we note that there are two ways to express the four-fermion expectation value $\langle c_\alpha^\dagger c_\beta c_\gamma^\dagger c_\delta \rangle$ in terms of $S_\bq$ 
\begin{align}
    &\langle c_\alpha^\dagger c_\beta c_\gamma^\dagger c_\delta \rangle = \sum_\bG \int \frac{d^2\bq}{2\pi N_\Phi}  \Lambda^\bq_{\alpha\beta} S_{\bq,\bG} \Lambda^{-\bq + \bG}_{\gamma\delta} \nonumber \\
    &= \delta_{\gamma, \delta} \langle c_\alpha^\dagger c_\beta \rangle + \delta_{\beta,\delta} \langle c_\alpha^\dagger c_\gamma \rangle - \langle c_\alpha^\dagger c_\delta c_\gamma^\dagger c_\beta \rangle \nonumber \\
    &= \delta_{\gamma \delta} \langle c_\alpha^\dagger c_\beta \rangle + \delta_{\beta,\delta} \langle c_\alpha^\dagger c_\gamma \rangle - \sum_\bG \int \frac{d^2\bq}{2\pi N_\Phi} \Lambda^\bq_{\alpha \delta} S_{\bq,\bG} \Lambda^{-\bq + \bG}_{\gamma \beta} 
\end{align}
Acting with $\Lambda_\bp^{\alpha\beta} \Lambda_{-\bp + \bG}^{\gamma \delta}$ and summing over $\alpha,\beta,\gamma,\delta$ yields
\begin{multline}
    S_{\bp,\bG} = -\sum_{\bG'} \int \frac{d^2 \bq}{2\pi} \Gamma^{\bq,\bG'}_{\bp,\bG} S_{\bq,\bG'} \\+ \tr \Lambda_\bp M \tr \Lambda_{-\bp + \bG} + \tr \Lambda_\bp \Lambda_{-\bp + \bG} M^T
    \label{Generalized_Haldane_Duality}
\end{multline}
where we define $\Gamma^{\bq,\bG'}_{\bp,\bG} := \frac{1}{N_\phi} \tr \Lambda^T_{\bp} \Lambda^\bq \Lambda^T_{-\bp+\bG} \Lambda^{-\bq + \bG'}$ and $M_{\alpha \beta} := \langle c_\alpha^\dagger c_\beta \rangle$. Note that in deriving (\ref{Generalized_Haldane_Duality}), we have only used the fermionic anticommutation relation and the invertibility of the form factors (\ref{invertability_form_factor}). 

 For the LLL, we can use $S_{\bq,\bG} = S_\bq \delta_{\bG,0}$ and $\Lambda_\bq = e^{-i \bq \cdot \bR}$. Furthermore, we can show that $\Lambda^\bq = \Lambda_\bq^* = \Lambda_{-\bq}^T$, $\Lambda_\bq \Lambda_\bp = e^{-\frac{i}{2} \bq \wedge \bp} \Lambda_{\bp + \bq}$ (the latter follows from the commutation relations of the guiding center $\bR$) and $M_{\alpha \beta} = \nu \delta_{\alpha, \beta}$. This yields $\Gamma^{\bq,0}_{\bp,0} = e^{i \bq \wedge \bp}$, $\tr \Lambda_{-\bp} = N_\Phi \delta_{\bp,0}$ and $\tr M = N$ which gives the Haldane duality (\ref{Haldane_duality}) upon substituting in (\ref{Generalized_Haldane_Duality}).
 
We note that the derivation above only used the properties $\Lambda^\bq = \Lambda_\bq^* = \Lambda_{-\bq}^T$, $\Lambda_\bq \Lambda_\bp = e^{-\frac{i}{2} \bq \wedge \bp} \Lambda_{\bp + \bq}$ which hold irrespective of the basis and gauge choice we use. For our explicit computations, we will use a Bloch basis for the lattice magnetic translation operators labelled by BZ momentum $\bk$ and assume a periodic gauge. This yields the explicit expression for $\Lambda_\bq$:
 \begin{equation}
     \Lambda_\bq^{\bk,\bk'} = e^{-\frac{i}{2} \bq \wedge \bk} \alpha_{\bk'}^* \alpha_{\bk+\bq} \delta_{[\bk'],[\bk + \bq]}
 \end{equation}
 where $[\bk]$ denotes the BZ part of $\bk$, $\{\bk\}:= \bk - [\bk]$ is the reciprocal lattice part and $\alpha_\bk = \eta_{\{\bk\}} e^{\frac{i}{2}[\bk] \wedge \{\bk\}}$. More details about the form factor, its inverse, and the Bloch basis are given in S.M.~\cite{SM}.

This duality plays a big role in fixing the anti-symmetrization (fermonic exchange statistics) of the 2RDM which can be seen from its derivation. Besides that, we want to point out several extra constraints that can also be inferred by the duality: the first one is the sum rule (equivalent to the trace condition in 2RDM):
\begin{equation}
    \int \frac{d^2\bq}{2\pi} S_\bq = N-N^2+NN_\Phi;
\end{equation}
and the second one is the large $\bq$ behavior of the $S_\bq$:
\begin{equation}
    \lim_{\bq\to\infty}S_\bq = N-N^2/N_\Phi.
\end{equation}
These two are obtained by taking $\bp=0$ and $\bp = \infty$ in Eq.~\eqref{Haldane_duality}, respectively. A third identity is obtained by multiplying both sides of Eq.~\eqref{Haldane_duality}
by $e^{-\frac{1}{2} \bp^2}$ and integrating over $\bp$ leading to
\begin{equation}
    \int \frac{d^2 \bq}{2\pi} S_\bq e^{-\frac{1}{2} \bq^2} = N
\end{equation}
One way to understand this relation is to note that the right hand side is nothing but (twice) the energy expectation value $E[S_\bq]$ (\ref{ESq}) for $V_\bq = 1$, which corresponds to a contact interaction. However, writing $e^{-\frac{1}{2} \bq^2} \rho_\bq \rho_{-\bq} = e^{-\frac{1}{2} \bq^2} :\rho_\bq \rho_{-\bq}: + \rho_0$, we note that the first (normal-ordered) part vanishes when acting on any fermionic state since contact interaction does nothing when acting on spinless fermions. Thus, for $V_\bq = 1$, $2E[S_\bq] = N$.

We note that the invertibility of the LLL form factor follows directly from the fact that the linear vector spaces spanned by $c_\alpha^\dagger c_\beta$ and $\rho_\bq$ have the same dimension $N_\Phi^2$. This condition is equivalent to the existence of an invertible linear map between the projected density operator in a band and a set of `unphysical' density operators satisfying the Girvin-MacDonald-Platzmann (GMP) algebra~\cite{GMP1986}. Such a map was first discussed by Shankar and Murthy \cite{ShankarMurthy} in the context of general Chern bands and was discussed by one of the authors~\cite{Jie_GMP} in the context of ideal bands. We expect this property to remain true for bands obtained by perturbing the LLL by a small amount. We leave the full characterization of the invertibility of form factors in general bands to future works. 

\subsection{Positivity constraints in terms of $S_q$}

In this section, we discuss how the general positivity constraints of the 2-RDM are expressed in terms of $S_\bq$ and discuss the simplifications in this representation, particularly by exploiting CMTS.

\subsubsection{$(2,2)$ constraints}

Let us start with the simplest $(2,2)$ positivities which contain three constraints. First, the G condition $\langle \O_G \O_G^\dagger \rangle \geq 0$ with $\O_G = \sum_{\alpha\beta}A^G_{\alpha\beta}c^\dagger_\alpha c_\beta$ implies the matrix $\mathcal G^{\alpha \beta}_{\delta \gamma} = \langle c_\alpha^\dagger c_\beta c^\dagger_\gamma c_\delta \rangle$ is positive semidefinite, $\G \succeq 0$. Using Eq.~\eqref{invertability_form_factor}, the matrix $\G$ can be expressed in terms of the structure factor $S_\bq$ via
\begin{equation}\label{G_condition_in_Sq}
    \G^{\alpha \beta}_{\delta \gamma}=  \int \frac{d^2 \boldsymbol q}{2\pi} \Lambda^{\boldsymbol q}_{\alpha \beta} \Lambda^{-\boldsymbol q}_{\gamma \delta} S_{\boldsymbol q}
\end{equation}
Noting that $\Lambda^{-\bq}_{\alpha \beta} = [\Lambda^\bq_{\beta \alpha}]^* = \Lambda_\bq^{\beta \alpha}/N_\Phi$ and recalling Eq.~\eqref{invertability_form_factor}, we see that Eq.~\eqref{G_condition_in_Sq} above implies that the form factor matrix $\Lambda$ diagonalizes $\G$ with the eigenvalues being the values of the structure factor $S_\bq$ at different $\bq$'s. Crucially, this means that the \emph{PSD} condition $\G \succeq 0$ reduces to the simple \emph{linear} condition $S_\bq \geq 0$ (see S.M.~\cite{SM} for a detailed discussion).

Next, we consider the D condition with operator $\O_D = \sum_{\alpha \beta} A^D_{\alpha \beta} c^\dagger_\alpha c^\dagger_\beta$, with $A^D_{\alpha \beta} = -A^D_{\beta \alpha}$, the positivity of $\langle \O_D \O_D^\dagger \rangle$ translates to the statement that the 2RDM $\D^{\alpha \beta}_{\gamma \delta} = \langle c_\alpha^\dagger c_\beta^\dagger c_\delta c_\gamma \rangle$ is PSD. This can be expressed in terms of the structure factor via
\begin{equation}
    \label{M2Constraint}
    \D^{\alpha \beta}_{\gamma \delta} = -\nu \delta_{\alpha \delta} \delta_{\beta \gamma} + \int \frac{d^2 \boldsymbol q}{2\pi} \Lambda^{\boldsymbol q}_{\beta \delta} \Lambda^{-\boldsymbol q}_{\alpha \gamma} S_{\boldsymbol q}, \quad \D \succeq 0
\end{equation}
The Haldane duality (\ref{Haldane_duality}) guarantees the antisymmetry of $\D^{\alpha \beta}_{\gamma \delta}$ under exchanging $\alpha \leftrightarrow \beta$ or $\gamma \leftrightarrow \delta$. A more physically suggestive way to write $\D$ is to express it as a first quantized 2-particle Hamiltonian:
\begin{equation}\label{effectiveHamiltonian}
    \D = \nu (1 + \nu) + \int \frac{d^2 \bq}{2\pi} s_\bq e^{i \bq \cdot (\bR_1 - \bR_2)}.
\end{equation}
This corresponds to the Hamiltonian of two particles in the LLL interacting via an effective Coulomb potential $\tilde V_\bq  = s_\bq e^{\frac{1}{2} \bq^2}$. Note that $\sum_{i,j = 1}^p e^{i \bq \cdot (\bR_i - \bR_j)} = p + \sum_{i \neq j = 1}^p e^{i \bq \cdot (\bR_i - \bR_j)}$ corresponds to the first quantized action of the guiding density operator $\rho_\bq$ on a wavefunction of $p$ particles. Then, the positive semi-definiteness of $\D$ translates to the statement that the spectrum of the corresponding effective Hamiltonian is non-negative. As we will show later in Sec.~\ref{feasible_region}, the spectrum of (\ref{effectiveHamiltonian}) admits a simple form whose energies are also linear functionals of $s_\bq$ provided that $s_\bq$ is rotationally symmetric. While this is generally not true on a finite torus, we expect it to be true either on geometries with rotation symmetry such as sphere or disk or for sufficiently large system size. Hence, while we will impose this constraint as a PSD constraint, we will see later in Sec.~\ref{feasible_region} that we can analyze its behavior as a linear constraint.

Finally, we can consider the Q condition which is particle-hole conjugated version of the D condition, $\langle \O_D^\dagger \O_D \rangle \geq 0$. It is straightforward to see (see S.M.~\cite{SM} for a detailed discussion) that the Q condition for quantum Hall is strictly weaker than the D condition for $\nu < 1/2$ and strictly stronger for $\nu > 1/2$ so it suffices to consider only one of the two.

\subsubsection{$(2,3)$-T1 \& T2 constraints}

We now consider the next level in the constraint hierarchy as discussed in Sec.~\ref{2_p_constraints}. As we discussed earlier, there are two constraints built from the identity $\sum_i \langle \O_i^\dagger \O_i \rangle \geq 0$. The first, called T1 constraint, is constructed by taking $\O_{T1,1} = \sum_{\alpha\beta\gamma}A_{\alpha\beta\gamma}c^\dagger_\alpha c^\dagger_\beta c^\dagger_\gamma $ and $\O_{T1,2} = \O_{T1,1}^\dagger = \sum_{\alpha\beta\gamma}A^*_{\alpha\beta\gamma}c_\gamma c_\beta c_\alpha$ whereas the second, called the generalized T2 constraint~\cite{Mazziotti2012}, is constructed by taking  $\O_{T2,1}\equiv\sum_{\alpha\beta\gamma}A_{\alpha\beta\gamma}c^\dagger_\alpha c^\dagger_\beta c_\gamma + \sum_\alpha B_\alpha c^\dagger_\alpha $ and $\O_{T2,2}\equiv\sum_{\alpha\beta\gamma}A^*_{\alpha\beta\gamma}c^\dagger_\gamma c_\beta c_\alpha   +\sum_\alpha D^*_\alpha c_\alpha$. In both cases, the expression $\sum_i \O_i^\dagger \O_i$ only contains terms of the form $c^\dagger c^\dagger c c$ and $c^\dagger c$ due to the cancellation of the term $c^\dagger c^\dagger c^\dagger c c c$. The term $\langle c^\dagger c^\dagger c c \rangle$ can be expressed as a linear functional of $S_\bq$ using (\ref{M2Constraint}) whereas $\langle c_\alpha^\dagger c_\beta \rangle = \nu \delta_{\alpha \beta}$ due to CMTS. Thus, both constraints take the form $\M[S_\bq] \succeq 0$ for some matrix $\M$ which depends linearly on $S_\bq$.

The (2,3) constraints can be simplified considerably using CMTS as follows. To do this, we will use a basis of Bloch states labeled by crystal momenta. Such a basis is constructed by choosing basis vectors enclosing a unit cell containing a $2\pi$ flux such that magnetic translation symmetries corresponding to the basis vectors commute \cite{SM}. This basis is natural when working on the torus geometry and has the advantage that the form factors are sparse matrices. To see how this basis simplifies the $(2,3)$ constraints, let's rewrite $\O_{T1,1} = \sum_{\bk_1,\bk_2,\bQ} A^{\bQ}_{\bk_1,\bk_2} c_{\bk_1} c_{\bk_2} c_{\bQ-\bk_1-\bk_2}$ and similarly for $\O_{T1,2}$. Then, we will find that $\sum_i \langle \O_{T1,i}^\dagger \O_{T1,i} \rangle = \sum_\bQ [A^\bQ_{\bk_1,\bk_2}]^* \M^\bQ_{\bk_1,\bk_2;\bk'_1,\bk'_2}[S_\bq] A^\bQ_{\bk'_1,\bk'_2}$ \cite{SM}. This reduces the PSD constraint $\M[S_\bq] \succeq 0$ where $\M$ has linear dimension $\sim N_\Phi^3$ to $N_\Phi$ PSD constraints $\M^\bQ[S_\bq] \succeq 0$ where the matrices $\M^\bQ$ are much smaller with  linear dimension going only as $\sim N_\Phi^2$. Note that this simplification only relies on lattice translation symmetry, which is applicable beyond the LL problem and is a subgroup of CMTS. As we show in SM \cite{SM}, the full CMTS allows us to relate the matrices $\M^\bQ$ for different $\bQ$s through unitary transformations which reduces the number of PSD constraints to $O(1)$ constraints \cite{SM}. Overall, this reduces the complexity of the T1/T2 constraints from imposing the PSD condition of $\sim O(1)$ matrices with linear dimensiotn $\sim N_\Phi^3$ to that of $\sim O(1)$ matrices of linear dimenstion $\sim N_\Phi^2$.

\subsection{Convex optimization}
On a practical level, one of the main advantages of the bootstrap and related methods is that they transform the problem into a convex optimization problem. Importantly, convex optimization is capable of solving problems with \emph{non-linear} PSD constraints through semidefinite programming (SDP). The latter problem is now reasonably well-studied with well-developed algorithms, especially the \textit{primal-dual interior point method}~\cite{wright1997primal,alizadeh1998primal} that we use in this work. The SDP implemented in this work is carried out using MATLAB~\cite{MATLAB} together with the MOSEK~\cite{mosek} solver. We also use the comprehension layers YALMIP~\cite{lofberg2012} and/or CVX~\cite{CVX,gb08} to convert our program into a standard form in SDP. 

We close Sec.~\ref{problem_setup} by summarizing the Bootstrap setup used in this work:
\begin{equation}
\begin{split}
    \textbf{Minimize } &E[S_\bq], \\
    \textit{subject to } &\text{Haldane duality [Eq.~\eqref{Haldane_duality}]}\\
    &+\text{ 2RDM positivity constraints.}
\end{split}
\end{equation}
Here the 2RDM constraints are implemented as linear matrix inequality (LMI) constraints having the form: $\sum_\bq S_\bq\Gamma^\bq-C\succeq 0$ where $\Gamma^\bq$'s and $C$ are given (pre-calculated) matrices and Haldane duality is an equality constraint. For example, for D condition, we have
\begin{equation}
    [\Gamma_D]^\bq_{\alpha\beta;\gamma\delta} = \frac{1}{N_\Phi}\Lambda^\bq_{\beta\delta}\Lambda^{-\bq}_{\alpha\gamma}\text{; } [C_D]_{\alpha\beta;\gamma\delta} = \nu \delta_{\alpha\delta}\delta_{\beta\gamma}.
\end{equation}
These can be easily formulated
through the comprehension layer and passed to the solver.

\section{Results on bootstrapping the lowest Landau level}
To obtain our results, we will use a screened Coulomb interaction $V_\bq = \frac{1}{\eta |\bq|} \tanh \eta |\bq|$ where $\eta$ controls the range of the interaction. This allows direct comparison with pseudopotential results where the $n$-th pseudopotential is given by 
$V_n = \int \frac{d^2 \bq}{2\pi} (V_\bq e^{-\bq^2/2}) L_n(\bq^2) e^{-\bq^2/2}$~\cite{Jie_Lattice}, where $V_\bq e^{-\bq^2/2} $ is the LLL-projected interaction.
We note that our density-density interaction differs from a normal-ordered interaction (which is the one used in first quantization for pseudopotential arguments) by the term $N E_0$ where $E_0 = \frac{1}{2} \int \frac{d^2 \bq}{2\pi} V_\bq e^{-\frac{\bq^2}{2}}$. In the following, we will measure the energies in units of $V_s$, the sum of all contributing pseudo-potential components:
$V_s\equiv\sum_{n=1,{\rm odd}} V_n$. This provides the natural energy scale for spinless fermions since all even terms drop out when acting on fermionic wavefunctions. It can be shown that the following is true in large system size with a full rotational symmetry:
\begin{equation}
    V_s = \frac{1}{4}\left( V_{\bq=0}-2E_0\right),
\end{equation}
which turns out to be a good measure even for small system sizes with no full rotational symmetry (i.e., system on torus) and will be used in our calculation.
In particular, this means that $V_s$ is parametrically smaller than $E_0$ for short-range interactions, $\eta \ll 1$, since $E_0(\eta\to 0) = V_{\bq=0}/2$.

Our results will be presented for two levels of constraints (2,2) and (2,2)+T1+T2. In this section (treating LLL), we find that imposing (2,2)+T2 ensures that T1 is already satisfied, so we will refer to the two levels of constraints as (2,2) and (2,2)+T2 or alternatively bootstrap without/with T2.

\begin{figure}[t]
    \centering
    \includegraphics[width=0.48\textwidth]{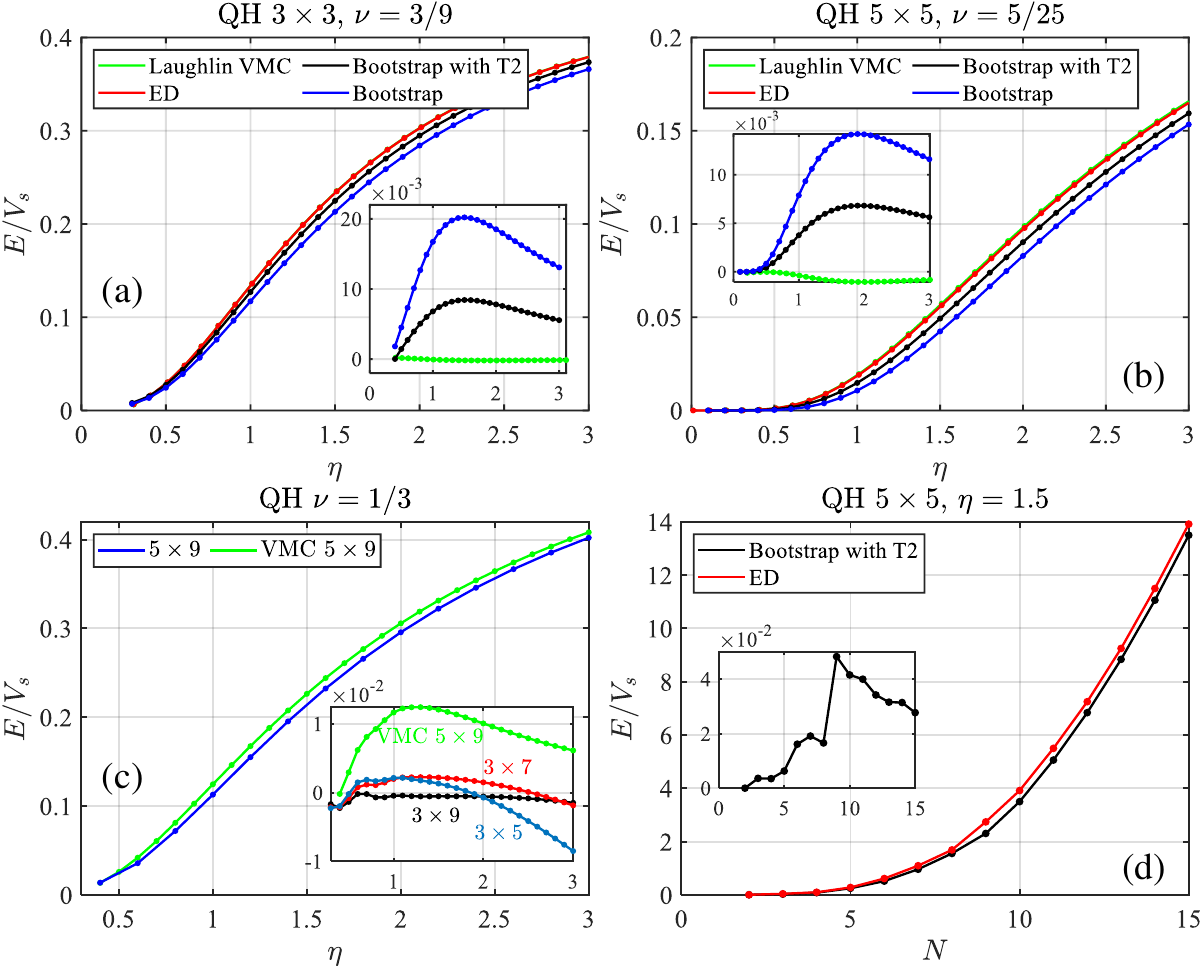}
    \caption{Energetics of the QH from Bootstrap, ED, and variational Monte Carlo (VMC). Energy per particle from Bootstrapping with only (2,2) (blue) and with (2,2)+T2 (black) constraints, from the ED (red), and from the VMC using Laughlin wavefunction on torus (green) for (a) $3\times3$ with $1/3$ filling and (b) $5\times5$ with $1/5$ filling, respectively, as functions of $\eta$. The inset shows the energy errors with respect to the ED data ($E_{\text{ED}}-E$). In panel (a), the VMC energies are so close to the ED so that they are totally covered by the red curve. (c) Energy per particle of Bootstrap with (2,2)+T2 for a large system size $5\times9$ at filling $\nu=1/3$ (blue) with VMC results (green) included as a comparison. The inset shows the relative energy of the results from other system sizes to the largest $5\times9$ grid ($E-E^{5\times9}_{\text{Bootstrap}}$) showing convergence of the results. $E^{5\times9}_\text{VMC}-E^{5\times9}_{\text{Bootstrap}}$ is also showed. (d) Total energy comparison between Bootstrapping with the (2,2)+T2 constraints (black) and the ED (red) for $5\times5$ as functions of the electron number $N$ at fixed $\eta=1.5$. The inset shows the energy error per particle with respect to the ED. All energies are measured in the unit of corresponding $V_s$, the sum of all contributing pseudo-potential components (see text).}\label{fig:BootstrapQH_energies_eta_filling_and_system_size_dependence}
\end{figure}

\subsection{Energetics: Comparison to ED and variational Monte Carlo (VMC)}
We start by comparing the energies we obtain in the bootstrap approach to Exact diagonalization (ED) at small system size, as summarized in in Fig.~\ref{fig:BootstrapQH_energies_eta_filling_and_system_size_dependence}. In panels (a) and (b), we compare the energies at fillings $\nu = 1/3$ and $1/5$ with ED on small system sizes $3 \times 3$ and $5 \times 5$, respectively. We see that our energy provides a lower bound to the actual energy as expected. For short range interaction, the error in the energy per particle is very small but increases as we increase the interaction range. With the $(2,2)$ constraint only, the maximum error in the energy per particle is around 1-2\% and goes down to 0.5-1\% when including the $(2,3)$ constraint. In panel (c), we show how our bound changes as we increase the system size indicating convergence as a function of system size. 
In addition to the comparison to ED, we also performed variational Monte Carlo (VMC) calculation since the Laughlin wavefunctions is expected to provide a very good variational wavefunctions for the $\nu = 1/3$ state that becomes exact in the limit of small $\eta$. Details about the model wavefunctions can be found in Appendix~\ref{model_wf}. The Laughlin VMC provides a rigorous variational upper bound for the ground state energy which can be combined with the bootstrap lower bounds to provide rigorous two-sided bounds on the energy. This simultaneously bounds the error for \emph{both approaches} and can be done for system sizes where ED is intractable. As shown in panel (c), we combine the VMC and bootstrap bounds to confine the exact ground state energy at filling $15/45$ within 1-2\% of the natural energy scale (see the green curve in the inset of panel (c)).

In panel (d), we show the energy obtained from the bootstrap vs ED for a $5 \times 5$ grid for different particle number $N$ with the inset showing the error per particle. The error per particle never exceeds $5\%$ suggesting our procedure captures the energetics reasonably well for all fillings. However, we see there is some systematic behavior in the error that can produce large errors when computing quantities that depend on the derivative of energy relative to particle number e.g. compressibility. In particular, there is a noticeable increase in the error on adding a particle to the $1/3$ Laughlin state (between $N = 8$ and $N = 9$) and a smaller one around the $1/5$ Laughlin state (between $N = 5$ and $N = 6$). This suggests the bootstrap approach underestimates the energy of quasielectron excitations on top of the Laughlin state (the error improves again as we add more particles). Later we will perform an in-depth analysis to identify the source of this error.

\subsection{$1/3$-filling}
\begin{figure}
    \centering
    \includegraphics[width=\linewidth]{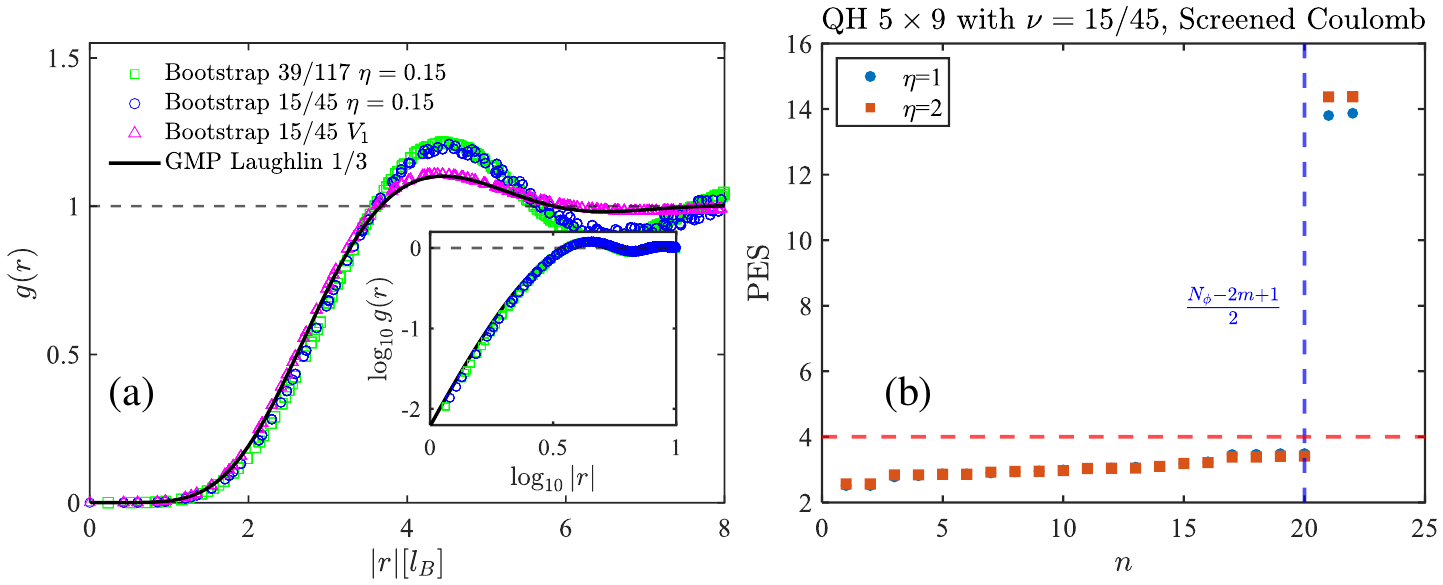}
    \caption{Pair correlation function $g(\br)$ and particle entanglement spectrum (PES) for 1/3 filling. (a) $g(r)$ from bootstrapping systems of filling $15/45$ with a screened Coulomb potential (blue circles) and the first pseudo-potential $V_1$ (purple triangles). The solid black curve indicates the results extracted from GMP~\cite{GMP1986}. The inset shows the log-log plot of the data. Large system calculation with filling $39/117$ with screened Coulomb is also shown (green squares). (b) PES from bootstrapping systems of filling $15/45$ with two different screening lengths: $\eta = 1$ (solid circle) and $\eta=2$ (square). The dashed lines indicate the gap in the PES, where the below gap count is shown explicitly which is $(N_\Phi-2m+1)/2$.}
    \label{fig:OneThird}
\end{figure}
To go beyond the energetics, we now look for specific correlation functions that can be constructed from the static structure factor $s_\bq$. For small screening distance $\eta$, we expect the ground state at fillings $1/m$, with $m$ odd and not very large, to be the Laughlin state. The main characteristic of the Laughlin state at filling $1/m$ is the existence of an order $m$ zero when any pair of particles approach each other. This is most readily detected from the pair correlation function $g(\br)$, which corresponds to the probability of having a pair of electrons separated by a distance $\br$. Its explicit form is given by $g(\br) = \langle c_0^\dagger c_\br^\dagger c_\br c_0 \rangle$, where $c_\br = \sum_\alpha \phi_\alpha(\br) c_\alpha$ and $\phi_\alpha(\br)$ is any basis of the LLL. It can be obtained from $s_\bq$ via the relation $g(\br)=\nu^{-1}-e^{-\br^2/2}-\nu^{-2}\int\frac{d^2 \bq}{2\pi}e^{-(\bq+\br)^2/2}s_\bq$. This expression can be obtained through applying Haldane duality to the expression in Ref.~\cite{GMP1986}. For the Laughlin state at filling $\nu = 1/m$, $g(\br)$ grows as $|\br|^{2m}$ at small $\br$ reflecting the order $m$ zero as the particles approach each other. In Fig.~\ref{fig:OneThird}a, we show the pair correlation function obtained in our approach for relatively short-range interaction, $\eta = 0.15$,  at $\nu = 1/3$. We see that our results perfectly reproduce the $|\br|^6$ power law at small distances. Comparing to the pair correlation function of the Laughlin state computed using variational Monte carlo \cite{GMP1986}, we see that while our approach reproduces most of the features of $g(\br)$, it  overestimates the oscillations of $g(\br)$ at large $\br$ and slightly underestimates the coefficient of the $|\br|^6$ term. 

Interestingly, when we take the interaction to be the first pseudopotential $V_1$, we reproduce the model state almost exactly at both small and large distances. In fact, our results show a discontinuity as a function of $\eta$: if the third pseudopotential $V_3(\eta)$ is smaller than the energy tolerance we obtain the Laughlin state almost exactly, but if $V_3(\eta)$ is above the tolerance then we obtain finite $\eta$ result depicted in Fig.~\ref{fig:OneThird}.
We will later discuss the source of this discontinuity.

While the pair correlation allows us to directly compare the correlation functions in our approach to those of the Laughlin state, it is important to consider other quantities that access more general aspects of the topologically ordered state that are not tied to a specific model wavefunction. This can be done by computing the entanglement spectra \cite{LiHaldane, OrbitalEntanglementBernevig, BulkEdgeEntanglement, ParticleEntanglementBernevig, ParticleEntanglementPapic}. Two different entanglement cuts are usually employed: a spatial/orbital cut \cite{LiHaldane, OrbitalEntanglementBernevig, BulkEdgeEntanglement} or a particle cut \cite{ParticleEntanglementBernevig, ParticleEntanglementPapic}. The former provides information on the edge excitations \cite{LiHaldane, OrbitalEntanglementBernevig, BulkEdgeEntanglement} whereas the latter provides information on the bulk quasiparticle excitations \cite{ParticleEntanglementBernevig, ParticleEntanglementPapic}.
The particle entanglement spectrum (PES) is directly related to the RDM. To see this, note that the $p$-PES is obtained by dividing the $N$ particles in the system into $p$ and $N - p$ particles, performing a Schmidt decomposition of the wavefunction and tracing over the $N-p$ part. The entanglement spectrum obtained is nothing but the negative logarithm of the eigenvalues of the $p$-RDM. Since our approach gives direct access to the 2RDM, we can obtain the 2-PES which is shown in Fig.~\ref{fig:OneThird}b. In general, the $p$-PES consists of a low energy sector separated by a large `entanglement gap' (which becomes infinite for model states) from high energy states. The number of low energy states is an indicator of the topological order and corresponds to the number of ways to place $p$ particles into $N_\Phi$ orbitals subject to so-called admissibility rules; for the $m$ Laughlin states, this simply means placing the particles on a ring with $N_\Phi$ sites such that any $m$ consecutive sites contain at most one particle. For $p = 2$, this gives $\frac{N_\Phi (N_\Phi - 2m + 1)}{2}$ low energy states. Since there are $N_\Phi$ distinct momenta and since the spectrum is momentum-independent due to CMTS, this leads to $\frac{N_\Phi - 2m + 1}{2}$ states when both $N_x$ and $N_y$ are odd. As we see in Fig.~\ref{fig:OneThird}b, the number of low energy states matches this count perfectly and the entanglement gap is relatively large even away from the small $\eta$ limit. We note here that the entanglement gap in the 2-PES is a measure of the smallness of the terms $|\br|^{2l}$, with $l < m$, in the expansion of $g(\br)$ at small distances. This means that it encodes the same information as $g(\br)$ but in a way that highlights the features associated with the topological order. 

\subsection{Half-filling}
\begin{figure}
    \centering
    \includegraphics[width=0.48\textwidth]{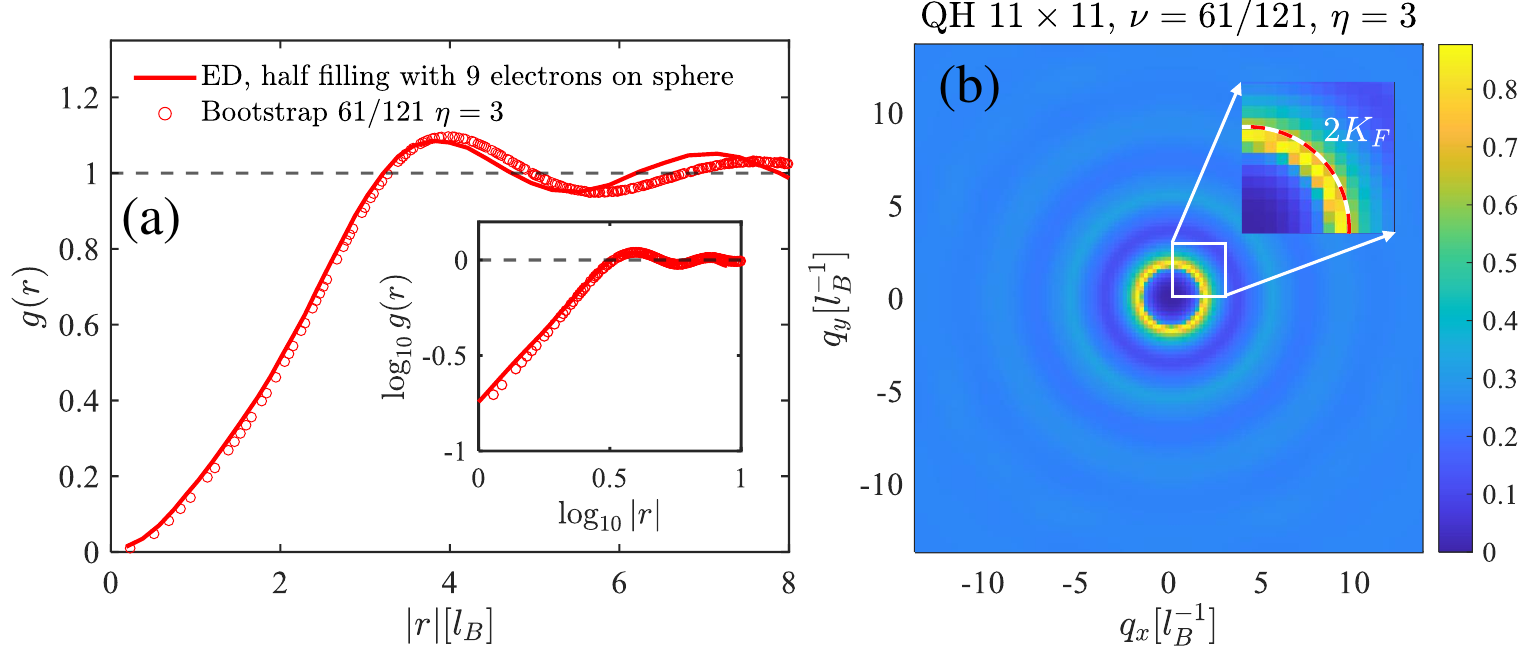}
    \caption{Pair correlation function $g(\br)$ and structure factor $s_\bq$ for half filling. (a) $g(\br)$ obtained from bootstrapping a system of filling $61/121$ with $\eta=3$ (red circle) and extracted from Ref.~\cite{RN413} for an ED calculation of half filling with 9 electrons on the sphere (red curve). The inset shows the log-log plot of the same data. (b) The 2D plot of the structure factor $s_\bq$ obtained from bootstrapping. The inset highlights the $2k_F$ feature.}
    \label{fig:half-filling}
\end{figure}
Next, we consider fillings close to $\nu = 1/2$ where the ground state is expected to be a compositie Fermi liquid (CFL)~\cite{Note6}. Unlike Laughlin states, the CFL state only has a first order zero when particles approach each other, which corresponds to $g(\br) \sim |\br|^2$.
We see in Fig.~\ref{fig:half-filling}a that our results not only reproduce this short distance asymptotic behavior, but also matches $g(\br)$ at all $\br$ when compared to ED data for a small system on the sphere \cite{RN413}. There is a slight mismatch in the oscillation period that may be due to different geometry and finite size effects. Remarkably, our approach seems to better reproduce the pair correlation function for the CFL at half-filling compared to the Laughlin state at $\nu = 1/3$. This is in contrast to most other methods such as mean field and DMRG which usually work better for gapped states. A more direct way to identify the CFL is a $2k_F$ feature in the static structure factor $s_\bq$, which is clearly identified in Fig.~\ref{fig:half-filling}b where there is a ring of peaks with the radius close to $2k_F$~\cite{Note7}. The ring structure also suggests that the rotation symmetry is reasonably good at relative small $\bq$.

To evaluate how tight our lower bound is, we can  compare the Bootstrap energy near half-filling to a VMC calculation using the CFL model wavefunction described in Appendix~\ref{model_wf}. One caveat in such comparison is that our bootstrap implementation assumes odd $N_\Phi$. There is no fundamental barrier to doing even $N_\Phi$, but it makes it easier to implement all symmetries and include the $\bq = 0$ point which we use to fix the particle number. To circumvent this, we perform the bootstrap calculation for fillings of $ (N_\Phi - 1)/2$ and  $ (N_\Phi + 1)/2$ electrons and interpolate between the two results. For sufficiently large odd $N_\Phi$, this should approach $\nu = 1/2$ result. For $N_\Phi=7\times9$ and $\eta=3$, we have $E^{\text{CFL}}_\text{bootstrap}/V_s\approx 0.789$ per particle while $E^{\text{CFL}}_\text{VMC}/V_s\approx 0.803$ per particle~\cite{Note8}. This corresponds to an energy error below 2\%. We reiterate that this number constitutes a rigorous bound on the sum of the errors of both VMC and the bootstrap.

\subsection{Characterization of the allowed region}\label{feasible_region}

To gain a deeper understanding of our results, we will now introduce a representation of the structure factor that will allow us to visualize the allowed region and the energy minimization procedure.
We note that for a geometry where rotation symmetry is a good symmetry, the structure factor $s_\bq$ only depends on $|\bq|$ and the Haldane duality implies that we can expand it in terms of odd Laguerre polynomials:
\begin{equation}\label{LaguerreExpansion}
    s_\bq = \nu (1 - \nu) + \sum_{n \text{ odd}} c_n L_n(\bq^2) e^{-\frac{\bq^2}{2}},
\end{equation}
where
\begin{equation}
    c_n = 2\int \frac{d^2\bq}{2\pi}\left( s_\bq - \nu (1 - \nu)\right)L_n(\bq^2) e^{-\frac{\bq^2}{2}}.
\end{equation}
Although we are working on the torus where full rotation symmetry is absent, we expect it to be recovered in the thermodynamic limit irrespective of geometry. The expansion~\eqref{LaguerreExpansion} should then yield a very good approximation. The approximate rotation symmetry of our results is also evident in Fig.~\ref{fig:half-filling}b. 

The advantage of this representation is that the effective Hamiltonian~\eqref{effectiveHamiltonian} describing the (2,2) constraint takes the simple form
\begin{equation}
    \D = \frac{1}{2}\sum_{n \text{ odd}} (4\nu^2 + c_n) P_n
\end{equation}
where $P_n$ is the projector on the space where the two particles have relative angular momentum $n$. Then, the PSD constraint $\D \succeq 0$ translates to the linear constraint $c_n \geq -4\nu^2$ (also linear in $S_\bq$ since the map between $c_n$ and $s_\bq-\nu(1-\nu)$ is linear). Furthermore, the energy takes the simple form 
\begin{equation}\label{eq:energy_laguerre_pseudo}
    E = \frac{N_\Phi}{2}\sum_{n \text{ odd}} (4\nu^2 + c_n) V_n
\end{equation}
where $V_n$ is the $n$-th pseudopotential component of the LLL-projected potential defined before. 

\begin{figure}
    \centering
    \includegraphics[width=\linewidth]{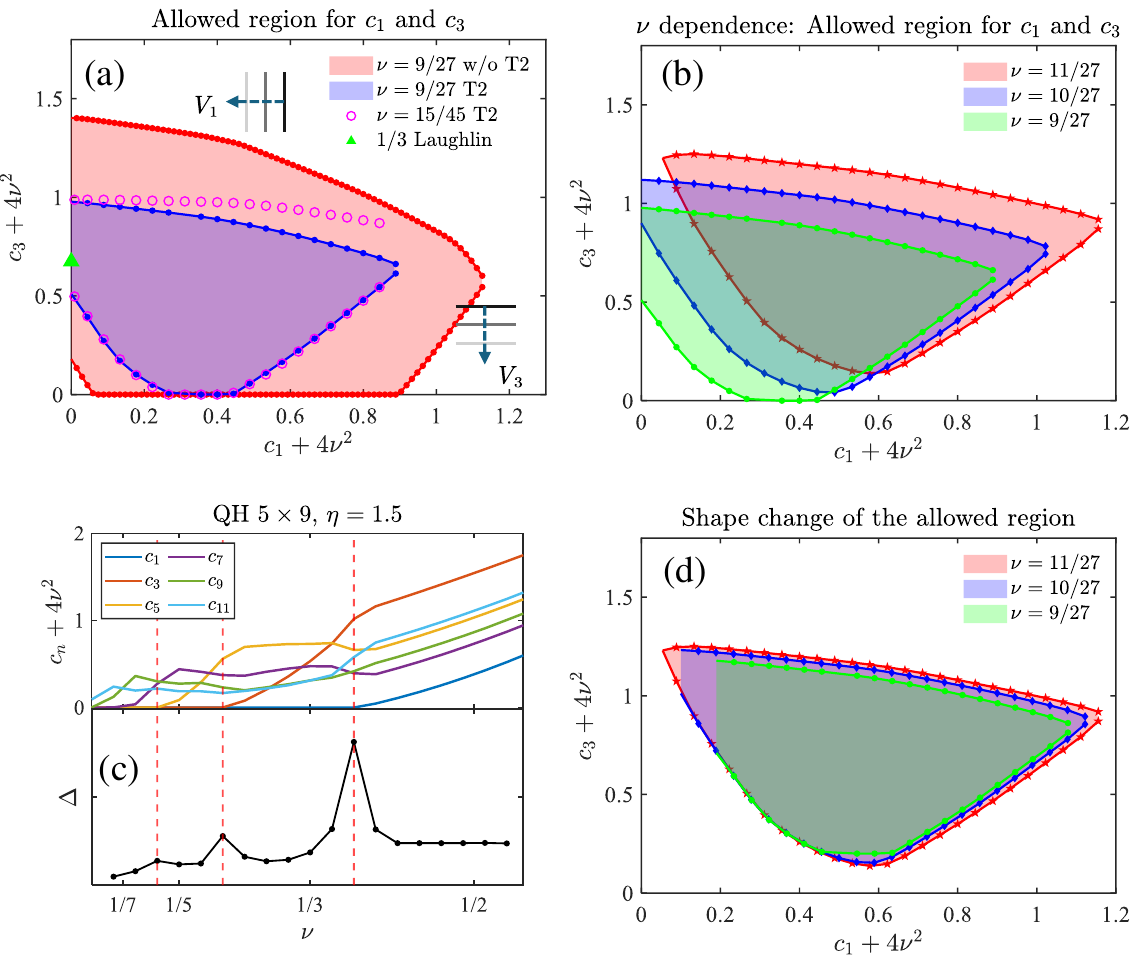}
    \caption{Allowed region characterized by the Laguerre polynomial basis \eqref{LaguerreExpansion},\eqref{eq:energy_laguerre_pseudo}. (a) Allowed region for $c_1$ and $c_3$ for a filling of $9/27$ obtained from imposing (2,2) (red) and (2,2)+T2 constraint (blue). The pink circles denote data from a larger system ($15/45$) calculation. The green triangle indicates the $1/3$ Laughlin state. (b) Filling dependence of the allowed region: $\nu=9/27$ (green), $\nu=10/27$ (blue), and $\nu=11/27$ (red). All data are obtained with T2 imposed. (c) Filling dependence of the coefficients ($c_1,c_3,\cdots,c_{11}$) and its relation to the charge gap $\Delta$. (d) Shifting the allowed region for different fillings in panel (c) so that they maximally overlap with each other. The green and blue regions are shifted by ($\Delta c_1 = 0.19,\Delta c_3 =0.2$) and ($\Delta c_1 =0.1,\Delta c_3 =0.11$), respectively, while the red region is not shifted.}
    \label{fig:AllowedRegion}
\end{figure}

Let us now understand our results in terms of the expansion coefficients $c_n$. For short-range interactions, we have $V_1 \gg V_3 \gg V_5 \dots$, thus energetically we are trying to get $c_1 + 4\nu^2$ as close to zero as possible followed by $c_3 + 4\nu^2$ and so on. The other constraints, $s_\bq \geq 0$ and $(2,3)$ constraints impose other non-trivial filling-dependent constraints on the coefficients $c_n$. In Fig.~\ref{fig:AllowedRegion}a, we map the allowed region at $\nu = 1/3$ in the $c_1 - c_3$ plane in the presence of only (2,2) or (2,3) constraints. We see that, as expected, the allowed region shrinks as we add more constraints. In the limit of short-range interaction, $V_3 \ll V_1$ and we should have $c_1 \approx -4\nu^2$. This should uniquely lead to the Laughlin state marked by the green triangle in Fig.~\ref{fig:AllowedRegion}a. Instead, since our constraints are not sufficiently tight, we get an extended region of allowed $c_3$ along the $y$-axis surrounding the Laughlin state. We note that the extended region does not extend all the way to the bottom of the $y$-axis indicating that our approach can detect the fact that adding a small $V_3$ always yields a nonzero energy. Since our approach always aims to minimize the energy within the feasible region, in the presence of any finite $V_3$ we get the smallest allowed value of $c_3 + 4\nu^2$. For our (2,2) and (2,3) constraints, this is around $0.5$, compared to $0.66$ for the Laughlin state. This discrepancy is the source of the error in the pair correlation function at large $\br$. To explain why we got something very close to the model state if we just use $V_1$ as the interaction, note that in this case, the energy is degenerate along the line $0.5 \lessapprox c_3 + 4\nu^2 \lessapprox 1$. In the case of degeneracy, the interior point method we use picks out the point furthest from the upper and lower boundaries, leading to a state very close to the Laughlin state~\cite{Note9}.

The error in the energy as a function of filling can also be understood within this picture. For fillings larger than $1/3$, we expect the allowed region to detach from the $y$-axis; adding an electron to the Laughlin state should cost finite energy even with only the first pseudopotential $V_1$. However, because our feasible region is slightly larger than the physical one, we find that an allowed solution with $c_1 = -4\nu^2$ persists for a range of filling after $1/3$ before getting detached. This can be seen in Fig.~\ref{fig:AllowedRegion}b, where we find a range of allowed solutions with $c_1 = -4\nu^2$ with $\nu = 10/27$ which only disappears at $\nu = 11/27$, i.e. after adding two electrons to the Laughlin state. We can also track this in Fig.~\ref{fig:AllowedRegion}c where we see the value of $c_1 + 4\nu^2$ taking off from 0 only at $\nu = 1/3 + \delta$ where $\delta \approx 0.05$. We can use the data for the energy as a function of $N$ to extract a charge gap $\Delta_N = E_{N+1} + E_{N-1} - 2E_N$, which is also equivalent to a discretized version of the inverse compressibility. We see that the shift $\delta$ mentioned above leads to a shift in the incompressibility peak expected at the Laughlin fractions as shown in Fig.~\ref{fig:AllowedRegion}c. The shift seems to be a constant for all Laughlin fractions, it scales with the total number of particles (so it should yield a finite shift in the filling in the thermodynamic limit) and is expected to get smaller as we add more constraints. 

The existence of this shift can be understood as follows. Our approach always tries to lower bound the energy subject to a set of constraints. Our set of constraints, constructed using few-body operators, likely accounts for few-body correlations. However, the fact that adding an electron to the Laughlin state costs finite energy arises due to global configurational constraints. This is evidenced by the difficulty of establishing a rigorous lower bound on the gap of such an excitation; proving that the Laughlin state has a charge gap even for pseudopotential Hamiltonians remains an open mathematical problem \cite{Rougerie2019laughlin} away from the so-called thin-torus limit \cite{Nachtergaele2021spectral}. It is also suggested by the construction of Laughlin states in second quantization \cite{ortiz_repulsive_2013, chen_algebraic_2015} or in terms of Jack polynomials \cite{bernevig_generalized_2008, bernevig_model_2008, bernevig_properties_2008}. Here, the energy of a state, while depending on pairwise interactions, is related to a clustering property in the occupation number basis which is restricted by global occupation number configurations in a non-trivial way. We expect such kinds of global filling-dependent configuration constraints to only be captured on average when we only account for few-particle correlations. An interesting future direction is whether such constraints can be phrased as PSD constraints that can be incorporated in the bootstrap approach.

Finally, we would also like to point out an interesting observation. The convergence of the boundary of the allowed region with system size is very different for the upper and lower boundaries (see pink circles in Fig.~\ref{fig:AllowedRegion}a). To understand this, recall that the lower boundary generally corresponds to ground states obtained by minimizing $V_3$ with positive coefficient whereas the upper boundary corresponds to minimizing $V_3$ with negative coefficient. The former is the standard quantum Hall problem with repulsive interaction where the convergence in system size indicates a short correlation length as expected. The latter corresponds to the LLL with \emph{attractive} interaction which likely stabilizes a different class of states. Thus, the results of Fig.~~\ref{fig:AllowedRegion} already provides some information about the behavior of a LLL with attractive short-range interaction. This observation underscores the power of the RDM bootstrap approach in extracting a broad range of information about the system.

\section{Diagnosing Phase transitions}
In this section, we will briefly discuss how to diagnose phase transitions within the bootstrap approach discussed in this work. We will leave a more detailed systematic investigation for future works. For simplicity, we will focus on first-order transitions although our analysis can be generalized to higher-order transitions as well. Let us say we have a first order transitions between two phases characterized by the 2-RDMs $\D_1$ and $\D_2$. At the transition point, the energies of $\D_1$ and $\D_2$ are the same which means that any 2-RDM $\D = \lambda \D_1 + (1 - \lambda) \D_2$ also has the same energy. Since the energy is a linear functional of the 2-RDM for any two-body interaction, this means that the boundary of the allowed region must have a straight line perpendicular to the direction defined by the Hamiltonian. This will lead to a kink in the energy as a function of the direction of the Hamiltonian (as we change the Hamiltonian parameters), reflecting a first order transition \cite{zauner2016symmetry}. Below, we will show how this manifests first for a simple toy example and then to study the transition from the CFL to possible non-Abelian states in higher LLs.

\subsection{A simple warm-up exercise}
\begin{figure}
    \centering
    \includegraphics[width=\linewidth]{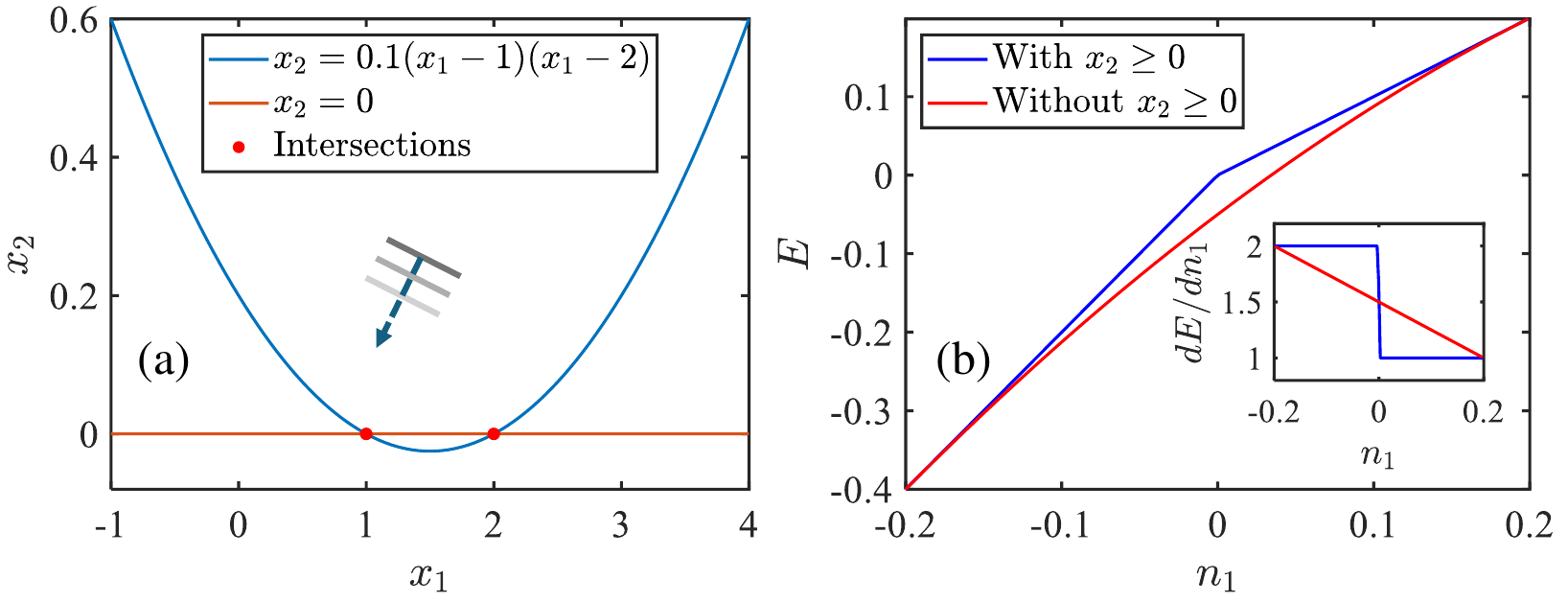}
    \caption{A simple toy model to illustrate the phase transition. (a) The boundaries of the two constraints in the optimization problem~\eqref{toy_model_opt} with the red dots highlighting the intersections of the two constraints. The arrow represents the descending direction of minimization process, which is $-\boldsymbol{n}$. (b) The optimized objective function values $E$ for two constraints (blue curve) and for only one constraint without the $x_2\geq 0$ (red curve) as functions of the parameter $n_1$ while $n_2$ is fixed to be 2. The inset shows the derivative of $E$ with respect to $n_1$ indicating a sudden jump in the blue curves.}
    \label{fig:toy_model}
\end{figure}
Consider the following optimization problem:
\begin{equation}\label{toy_model_opt}
\begin{split}
    \textbf{Minimize } &E = n_1x_1+n_2x_2, \\
    \textit{subject to } &x_2\geq 0.1(x_1-1)(x_1-2)\text{ and } x_2\geq 0
\end{split}
\end{equation}
where $x_1$ and $x_2$ are the optimization variables and we have two constraints: one is non-linear and the other is linear. The vector $\boldsymbol{n}\equiv(n_1,n_2)$ defines the direction of the optimization. A visualization of this optimization problem is given in Fig.~\ref{fig:toy_model}(a). 

Without even using any solver, we see that there is a jump in the optimized $x_1$ values (denoted as $x_1^o$) when the vector $\boldsymbol{n}$ is near $(0,n_2>0)$: for $\boldsymbol{n}=(0.01,2)$, $x_1^o=1$; for $\boldsymbol{n}=(-0.01,2)$, $x_1^o=2$, as shown by two red dots in Fig.~\ref{fig:toy_model}(a). If one thinks of $x_1>1.5$ and $x_1<1.5$ as different phases, then we have a first-order phase transition (or first-order phase transition)%\PL{how about ``then we have a first-order phase transition" (remove ``sharp" which is not consistently defined)}
near the critical point $\boldsymbol{n}_c=(0,n_2>0)$. It is not hard to see that the constraint $x_2\geq0$ is what makes this sharp transition happen, and the vector $\boldsymbol{n}_c$ is precisely the normal vector to the boundary of that constraint. Thus, we define $x_2>0$ as the \textit{critical constraint} of the phase transition.

However, it is usually hard to visualize the allowed region in more complicated problems given its high-dimensional nature. Instead, we can see that the form of the allowed region given in Fig.~\ref{fig:toy_model}(a) immediately gives rise to a kink in the optimized energy (or a discontinuity in its first derivative) as a function of $n_1$  while fixing $n_2=2$ as shown in Fig.~\ref{fig:toy_model}(b). One way to identify the critical constraint responsible for this kink is to compare the optimized energy in the presence and absence of this constraint. As we see in the figure, switching off the $x_2\geq 0$ constraint (red curve) leads to the disappearance of the kink, suggesting it plays the role of the active constraint in perfect agreement with the visualization in panel a. We emphasize that this ability to switch constraints on and off is unique to the bootstrap approach and allows us to identify the type of correlations that are relevant close to the transition. 

\subsection{Landau level interpolation (LLI) and three-body correlations}
\begin{figure}
    \centering
    \includegraphics[width=\linewidth]{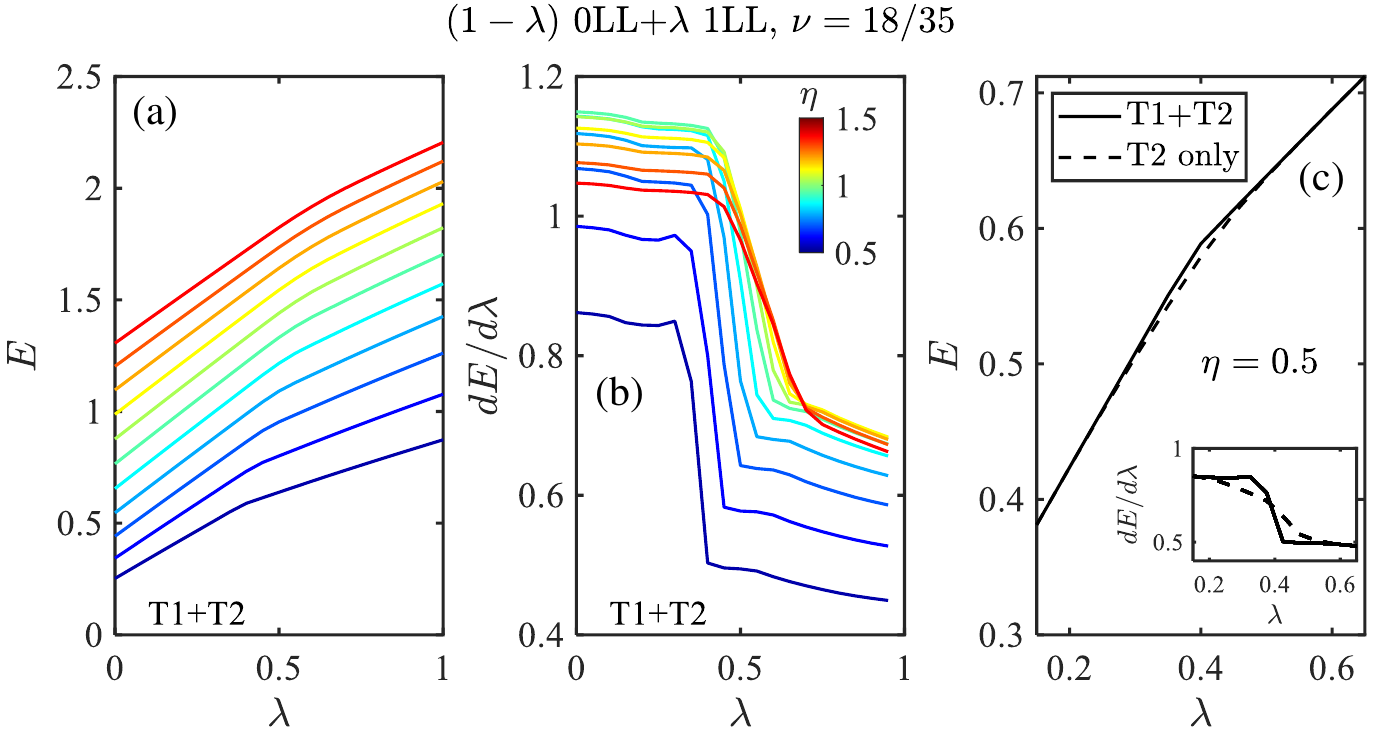}
    \caption{Phase transition between Landau levels near half-filling. (a) The optimized energies of the mixed Landau level system as a function of the mixing parameter $\lambda$. Curves with different colors represent different $\eta$'s in the potential profile, with the color bar shown in panel (b). (b) The corresponding derivatives of the energies with respect to $\lambda$. Results in (a,b) are obtained with T1 and T2 constraints. (c) A comparison of energies and their derivatives (inset) between T1+T2 (solid curve) and T2 only (dashed curve) optimizations. $\eta$ is fixed to be 0.5.}
    \label{fig:LLM_transition}
\end{figure}
As an application for the phase transition discussion above, we will now attempt to study the transition from CFL in the LLL at half-filling as we change the form factors to those of the first LL. For unscreened Coulomb interaction, we expect a transition to a non-Abelian gapped state such as Moore-Read \cite{mooreNonabelionsFractionalQuantum1991,bonesteelSingularPairBreaking1999,greiterPairedHallStates1992a,metlitskiCooperPairingNonFermi2015,papicQuantumPhaseTransitions2012,readPairedStatesFermions2000a,rezayi2000incompressible,wangPairingHalffilledLandau2014}. To this end, we consider the interaction
\begin{equation}
\begin{split}
    &V_{\rm LLI}(\bq,\lambda,\eta) \\
    &= \left[(1-\lambda)[L_0(\bq^2/2)]^2+\lambda[L_1(\bq^2/2)]^2\right]\frac{\tanh{\eta|\bq|}}{\eta|\bq|},
    \end{split}
\end{equation}
where the parameter $\lambda$ interpolates between the zeroth LL for $\lambda = 0$ and the first LL for $\lambda=1$. We note that such interaction can be realized in Bernal bilayer graphene where the wavefunctions of the lowest Landau levels are mixtures of 0th and 1st LL wavefunctions controlled by the bernal interlayer hopping parameter 
\cite{apalkovStablePfaffianState2011a,papicTunableInteractionsPhase2011,papicTunableElectronInteractions2011,papicTopologicalPhasesZeroth2014,zibrovTunableInteractingComposite2017,zhuWidelyTunableQuantum2020}. The parameter $\eta$ controls the interaction range as usual.

The results are shown in Fig.~\ref{fig:LLM_transition} for a system near half-filling, showing a clear kink in the energy curve as a function of the parameter $\lambda$ (see Fig.~\ref{fig:LLM_transition}(a)) for a wide range of $\eta$'s. This is also corroborated by calculating the first derivative of the energy as a function of $\lambda$ which displays a sudden jump (see Fig.~\ref{fig:LLM_transition}(b)). To identify the constraint responsible for the kink, we note that by switching off the T1 constraints, the kink is gone and we obtain a smooth energy curve instead Fig.~\ref{fig:LLM_transition}(c). The similarity to the toy model Fig.~\ref{fig:toy_model}(b) strongly suggests T1 has a similar role, becoming important close to the transition point from the zeroth to the first Landau levels.

To understand the significance of the T1 constraint, recall it corresponds to the positivity of the expectation value $\langle c^\dagger c^\dagger c^\dagger ccc + ccc c^\dagger c^\dagger c^\dagger \rangle$ associated with three-body correlations. At half-filling, the two three-body terms are related by particle-hole symmetry and this correlation function reduces to the 3-RDM. The T1 constraint becoming critical indicates that the 3-RDM starts developing zeros as we move from the zeroth to the 1st LL. Zeros of the 3-RDM indicate the existence of a three-body interaction that annihilates the corresponding state, i.e. we are transitioning to a state that is the ground state of a 3-body pseudopotential. This is consistent with the expectation of a non-Abelian (Moore-Read or Pfaffian state) as we move towards the first LL. Distinguishing the different non-Abelian topological states will likely require computing more complicated correlation functions and will be left to future works. We emphasize that, rather remarkably, our approach is able to diagnose the emergence of three-body correlations although our interaction is purely two-body. 

\begin{table*}[t]
\caption{Computational Complexity Overview [(2,2)+T2]}
\label{tab:complexity}
\begin{ruledtabular}
\begin{tabular}{ccccc}
\textrm{System Size} & \textrm{Symmetry Reduction}$^{\color{blue}\Box}$ & \textrm{CPU Hours}$^{\color{blue}\#}$ & \textrm{Memory (GB)} & \textrm{Solver}\\
\hline
$N_\Phi=3\times3$ & No  & $<0.1$   & $<1$   & MOSEK 10.1 \\
$N_\Phi=5\times 9$ & ${C_4}^{\color{blue}\dagger}$ & 100  & 200   & MOSEK 10.1 \\
$N_\Phi=7\times 9$ & $C_4$+Mirror$^{\color{blue}*}$   & 200 & 500  & MOSEK 10.1  \\
$N_\Phi=9\times 9$ & $C_4$+Mirror   & 800 & 1500  & MOSEK 10.1  \\
$N_\Phi=9\times 9$ & $C_4$+Mirror   & 800 & $<1500$  & COPT 7.2.4$^{\color{blue}\circ}$ \\
$N_\Phi=13\times 9$ & Full rotation + Sampling$^{\color{blue}\triangle}$   & $\sim1000$ & $<2000$ & MOSEK 10.1 \\
$N_\Phi=11\times 11$ & Full rotation + Sampling   & $\sim1000$ & $<2000$ & MOSEK 10.1 \\
\end{tabular}
\end{ruledtabular}
% \vspace{2pt}
\noindent
\footnotesize
\raggedright
${\color{blue}\Box}$: We have assumed CMTS for all simulations, so the listed reductions come from extra symmetries.
${\color{blue}\dagger}$: $S_\bq$ is constructed only in one of the $C_4$ sectors of the $\bq$ lattice ($N_\Phi\times N_\Phi$).
${\color{blue}*}$: 4 mirror axes: $q_{x,y}=0$ and $q_x = \pm q_y$.
${\color{blue}\triangle}$: Details can be found in Sec.~\ref{sampling}.
${\color{blue}\circ}$: We include the state-of-art SDP solver COPT~\cite{copt} for comparison which performs better than Mosek. 
${\color{blue}\#}$: Intel Sapphire Rapids 112-core CPU\\
% In most of the test, it performs much better than Mosek showing great potential of optimization on the solver side.\\
\textbf{Note:} The results are also very sensitive to the settings of the solver such as tolerance. We used default settings. All values of CPU hours and memory usage are roughly estimated since it is hard to get exact values from the MATLAB interface. The sizes of $9\times9$ and $11\times 11$ are done for half-filling while all others are done for one third filling. The complexity of this algorithm doesn't depend on the filling $\nu$ or electron number $N$ meaning that different fillings will not change the order of run time and memory usage. For a reference, the $11\times 11$ case was running on the cluster (one node with 112-core CPU) for roughly one day.
\end{table*}
\section{Computational analysis}
Now, let us move on to the numerical part and discuss in detail the complexity of using Bootstrap to deal with the quantum Hall problem. The complexity is essentially divided into two parts: the optimization problem itself (quantum Hall in our case) and the solver (algorithms used to solve the optimization problem). Thus, there are two parallel routes to improve efficiency: a better representation of the problem and a more robust and advanced solver. We will mainly focus on the first route, but will have a few comments regarding the second.
In this section, we provide a detailed complexity analysis with actual running hours. We also introduce a physics-inspired simplification that allowed us to go to very large system sizes that are comparable to state-of-the-art sizes accessible to other methods such as DMRG.

\subsection{Run time complexity}
It should be noted that although a polynomial-time algorithm means a problem can be efficiently solved in principle, scaling as a polynomial with a high power or large prefactor can still make a problem intractable in practice (though in this case, we may hope that some tweaks on optimizations can go a long way into making it tractable). Thus, techniques like symmetry reductions that can reduce the power or prefactor of the computational scaling become extremely important. In the following complexity analysis, we will primarily consider the T2 constraints. Other constraints can be analyzed similarly.

With CMTS encoded and using the $S_\bq$ representation, the SDP corresponding to T2 has the form of linear matrix inequality (LMI) with $N_\Phi^2$ variables (number of elements in $S_\bq$) and a matrix dimension of $(N_\Phi-1)N_\Phi/2\times (N_\Phi-1)N_\Phi/2$. From the standard convex optimization theory~\cite{vandenberghe1996semidefinite,boyd2004convex}, the time complexity using the primal-dual interior-point method is given by~\cite{alizadeh1998primal}
\begin{equation}
    O\left((m^3+m^2n^2+mn^3)\sqrt{n}\log(1/\epsilon)\right),
\end{equation}
where $m=N_\Phi^2$, $n=(N_\Phi-1)N_\Phi/2$ and $\epsilon$ is tolerance for the duality gap. So, without any further optimization, bootstrapping the QH with the T2 constraint is roughly $O(N_\Phi^9)$ hard~\cite{Note10}. This allows us to do $N_\Phi\sim$ 10s which can take us beyond ED, but makes $N_\Phi\sim$ 100 difficult to realize.
The first and simplest optimization we did is to use the discrete symmetries to reduce the number of variables, i.e, the number independent elements in $S_\bq$. For example, since the $\bq$ lattice is alway square, we can use the $C_4$ symmetry to reduce the number of variables by a factor 4. Further considering the mirror symmetry with mirror axes $q_{x,y}=0$ and $q_x=\pm q_y$, we can reduce the number by a factor 8. Here we emphasize that this mirror isn't a symmetry of the fractional quantum Hall state (magnetic field chooses a chirality) but a symmetry of the structure factor $S_\bq=S_{|\bq|}$ since the mirror actions preserve the norm $|\bq|$~\cite{Note11}. The results are summarized in Table~\ref{tab:complexity}. One should notice that the particle number $N$ doesn't enter the complexity at all~\cite{Note12}, so we did not include the information about the particle number in the table as it doesn't change the run time too much.

\subsection{$q$-lattice sampling using approximated full rotational symmetry}\label{sampling}
The discrete symmetries discussed above only give us a constant reduction which does not change the scaling, meaning that the complexity is still $O(N_\Phi^9)$ only now with a much smaller prefactor. On the other hand, to enable large-system calculation, we can employ other  continuous symmetry of the quantum Hall problem as discussed below. Apart from CMTS, quantum Hall systems also have continuous rotation symmetry. This symmetry is broken on the torus but we expect it to be recovered in the limit of a sufficiently large torus. To account for this, we introduce the sampling method explained below.

Let us consider the structure factor $S_\bq$ on a $\bq$ lattice spanned by $\{ -q_{x}^{max},-q_{x}^{max}+\Delta q_x,\cdots,q_{x}^{max} \}$ and $\{ -q_{y}^{max},-q_{y}^{max}+\Delta q_y,\cdots,q_{y}^{max} \}$. We want to sample $\kappa$ points from a line segment: $\{ q_1=0,q_2,q_3,\cdots,q_{\kappa-1},q_\kappa = \sqrt{(q_{x}^{max})^2+(q_{y}^{max})^2} \}$ with $q_i$ ordered. We introduce a set of new optimization variables $\boldsymbol{S} = (S_1,S_2,\cdots,S_\kappa)$ such that the original $S_\bq$ can all be linearly interpolated from the new variables:
\begin{equation}
    S_{\bq} = \frac{|\bq|-q_{l}}{q_{r}-q_{l}} S_{l} + \frac{q_{r}-|\bq|}{q_{r}-q_{l}} S_{r},
\end{equation}
where $r,l\in\{1,2,\cdots,\kappa\}$ and $q_l\leq|\bq|\leq q_r$ with $q_{l,r}$ the two nearest neighbors of $|\bq|$. In doing so, the number of variables is reduced from $N_\Phi^2$ to $\kappa$ which is practically less than $N_\Phi$. Thus, the complexity is also reduced to $O(N_\Phi^8)$.

One might be worried about the inaccuracy induced by this simplification, since it reliess on the full rotational symmetry and $S_{|\bq|}$ being smooth. In fact, those two requirements are both approximately satisfied in quantum Hall on torus when the system size is sufficiently large. We tested this method with a uniform sampling of $N_\Phi$ points for a few system sizes. The result looks almost identical to the one using $C_4$ only with a difference of less than $10^{-4}$ in energy per particle. This allows us to perform very large system calculations as shown in Table~\ref{tab:complexity}. For example, the system size of $N_\Phi = 11\times 11$ is translated to a circumference of $L_y = 11\times \sqrt{2\pi} \approx 27$ in DMRG, which is comparable to state-of-the-art sizes achievable.

One final remark is that $S_{\bq}$ has a fast-damping tail ($S_{\bq\to\infty}=S_\infty$). Furthermore, for any physical quantity, the large-$\bq$ behavior is suppressed by the Gaussian factor $e^{-q^2/2}$. Thus, when sampling the $S_\kappa$, we can use less points for large $|q|$ without introducing any error in any physical observable such as energy or pair correlation function. This means, for very large systems, most of the $\bq$ points are irrelevant with only few small $\bq$'s needed. We call this the \textit{importance sampling}. With this, $m$ can be further reduced from $N_\Phi^2$ to just $O(1)$ (a constant) leading to an $O(N_\Phi^7)$ time complexity.

\subsection{SDP solvers}
The above analysis is based on the primal-dual interior-point method~\cite{alizadeh1998primal} which is a second-order algorithm and used in most of the modern SDP solvers like MOSEK~\cite{mosek} or COPT~\cite{copt} since it is the most robust algorithm applicable to all kinds of SDPs. However, it doesn't necessarily mean that it is the best algorithm for quantum many-body bootstrap using the RDM theory. In fact, it has been demonstrated that a first-order algorithm called boundary-point method can improve the efficiency by at least one order-of-magnitude for RDM calculations in quantum chemistry comparing to the interior-point method~\cite{mazziotti2011large}. We believe that developing a special solver for quantum many-body bootstrap will enable a similar speedup, that will enable us to access even larger system sizes and impose more constraints.

\section{Discussion and Outlook}

\subsection{Nature of correlation functions}
We begin by discussing a few subtleties and caveats regarding this method, particularly when used to compute ground state correlation functions (static structure factor in our case). While conventional variational method search a \emph{subset} of valid physical states and thus always outputs a physically realizable correlation function, even if it is not the correct one for the ground state, this method searches a \emph{superset} of valid physical correlation functions and thus its output is not always associated with any physical state. However, by enforcing enough constraints we can make sure these correlation functions are not easily distinguishable from physical ones, since if there was a computationally simple way to check that the obtained correlation is unphysical, we can add it to our set of constraints (at least if it has PSD form). This suggests that the more constraints we include, the harder it will be to distinguish the results we get from physical correlation functions. Our results on the quantum Hall problem suggest that our structure factors approximate the physical ones to a good degree. We leave a more rigorous discussion of how the output of this method approaches physical correlation functions to future works.

One feature we may expect in the output of this method is the following. If there are two closely competing states A and B, we may get some `compromise' correlation functions that combine energetically favorable features of both states, even if there is no physical state with both features. As the constraints get tighter, the method will be forced to choose the correlation function corresponding to one of the states. This may help explain some of the features of our results; the extra fluctuations at large $\br$ in $g(\br)$ at $\nu=1/3$ (Fig.~\ref{fig:OneThird}) may be associated with correlations of a nearby CFL or a Wigner crystal.
This is perhaps because the $\nu = \frac{1}{4}$ CFL optimizes $V_3$ in the degenerate space annihilated by $V_1$, and the bootstrap constraints are not yet tight enough to select out Laughlin as the unique zero mode of $V_1$ at $\nu = \frac{1}{3}$. Another option is a competitor Wigner crystal state, which in reality cannot be a zero mode of $V_1$ like the state in the bootstrap~\cite{Note13}. This picture suggests that constraints which rule out such compromise correlation functions between closely competing states can help significantly improve the results.

\subsection{Evolution of the allowed region}
\begin{figure}
    \centering
    \includegraphics[width=\linewidth]{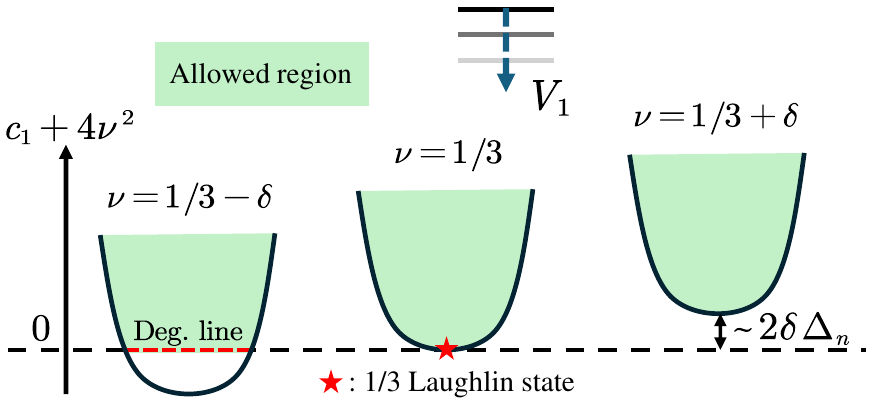}
    \caption{Ideal allowed region for LLL near $\nu=1/3$. The black dashed line represents the rigid wall ($c_1+4\nu^2=0$). The allowed regions for three consecutive fillings are shown: $\nu=1/3-\delta$ (left) which is cut through by the wall leading to a degenerate line (red dashed line) for the $V_1$ potential, $\nu=1/3$ (middle) touching the rigid wall with the $1/3$ Laughlin state sitting in the touching space (red star), and $\nu=1/3+\delta$ (right) leaving the hard with a gap proportional to the change of the filling: $2\delta \Delta_1$ where $\Delta_1$ is the charge gap of the pseudo-potential $V_1$ right above the $1/3$ filling.}
    \label{fig:shift_allowed_region}
\end{figure}
We now discuss the shift of the allowed region of the $c_n$. 
We comment on a curious observation. Apart from the rigid `walls' at $c_n + 4\nu^2 = 0$, the allowed region seems to evolve smoothly with filling, shifting with an almost constant vector with its shape change very little. This is illustrated in Fig.~\ref{fig:AllowedRegion}d, where we undo such shift showing the allowed regions for $\nu = 9/27, 10/27, 11/27$ almost coinciding away from the regions where they intersect the rigid walls as $c_n + 4\nu^2 = 0$. In fact, assuming the actual physically allowed region (i.e. the physical set of $N$-representable 2-RDMs, $\leftindex^2 D \in P^2_N$) also rigidly shifts with filling without significantly changing its shape, we can extract important physical information from the shift vector $\Delta \boldsymbol{c} = (\Delta c_1, \Delta c_3,\dots)$. We know that at fillings $\nu = 1/(n+2)$, for odd $n$, the physically allowed region has to intersect with the wall $c_{n'} = -4\nu^2$ for all $n'\leq n$. When $n=1$, namely the $1/3$ filling, such intersection becomes a touching point illustrated in Fig.~\ref{fig:shift_allowed_region} since the Laughlin state is the unique ground CMTS-symmetric ground state for $V_1$. For fillings below $\nu = 1/3$, we expect a degenerate line of zero and for fillings larger than $1/3$ we get a gap related to the shift. The shift vector near the filling $1/3$ (in our case, from $9/27$ to $10/27$) obtained from the data in Fig.~\ref{fig:AllowedRegion} has $\Delta c_1 = 0.09$ which corresponds to a gap $\Delta_1=\Delta c_1/(10/27-9/27)/2=1.215$, reasonably close to the gap computed in ED ($1.037$). This error can be further reduced when more constraints are included. Therefore, we believe this bootstrap picture provides a unique point of view for the pseudo-potential argument for quantum Hall. The discussion about other fillings for $n>1$ is more complicated than $n=1$ which will be discussed elsewhere.

\subsection{Further extensions: 3RDM}
\begin{figure}
    \centering
    \includegraphics[width=0.85\linewidth]{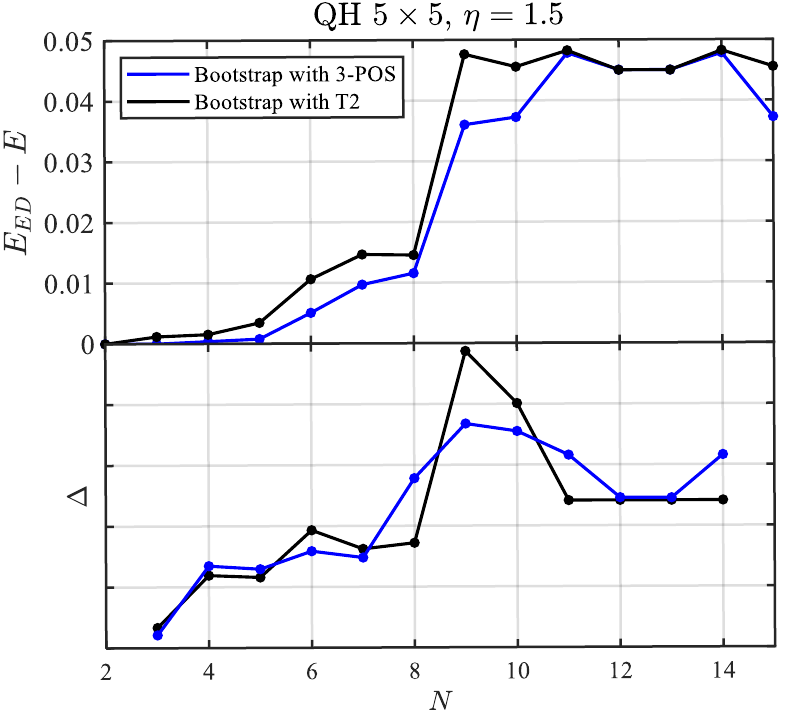}
    \caption{Comparison of results obtained from bootstrapping with (2,2)+T2 conditions (black) and with full (3,3) constraints (blue). Upper panel: the total energy error (not per particle as shown in the inset in Fig.~\ref{fig:BootstrapQH_energies_eta_filling_and_system_size_dependence}d) with respect to ED. Lower panel: the calculated charge gaps (or incompressibility).}
    \label{fig:3POS}
\end{figure}

One limitation of the 2-RDM approach discussed here is that constraints beyond (2,2), T1 and Generalized T2 requite tensor factorization of the variables and thus do not have the form of PSD constraints.
While there are possible approaches to deal with this limitation, which will be the topic for follow-up work, a straightforward way to go beyond our results is to consider higher RDMs. In this approach, instead of combining higher-body operators to produce a 2-body operator, we directly impose positivity constraints on the higher-body operators. Clearly, this approach is more memory intensive; 3RDM positivity constraints (namely, the (3,3) constraints) involve $O(N_\Phi^4)$ after accounting for CMTS. In fact, due to fermionic antisymmetry, the number of optimization variables is in practice $\sim 0.03N_\Phi^4$, making it achievable for reasonable system sizes. 

While an elaborate discussion about the higher RDM constraints will be given elsewhere, we provide the energy comparison in Fig.~\ref{fig:3POS}, which shows modest improvement compared to the T1/T2 constraints used in this work. The improvement in the relative error is largest for small fillings. Interestingly, our (2,3) results approach the full (3,3) for fillings close to $\nu = 1/2$. This can be understood by noting that particle-hole symmetry makes $\langle c^\dagger c^\dagger c^\dagger ccc\rangle$ equivalent to $\langle ccc c^\dagger c^\dagger c^\dagger\rangle$ at exactly half filling, thus indicating that the (2,3)-T1 and T2 combined are equivalent to the full (3,3) constraints. 
Note however that the 3RDM positivity constraints are still not sufficiently strong to fix the shift of the charge gap (as shown in the lower panel in Fig.~\ref{fig:3POS}), suggesting the need of different constraints as we argued earlier.

It is worth mentioning that it is possible to formulate a dual version of the RDM optimization \cite{Mazziotti2020, Mazziotti2023}, where the constraints become the optimization variables and RDM plays the role of a Lagrange multiplier. The main simplification in this formulation relies on the observation that for any solution of the optimization problem, only a small number of constraints is saturated or ``active'' (these will correspond to zero eigenvalues of the PSD constraint matrices). Furthermore, since the solution always lies on the boundary of the feasible region, there is a simple geometric meaning for those active constraints e.g. if there is only one active constraint, it has to describe the plane tangent to the feasible region which is perpendicular to the direction of the Hamiltonian. We leave a detailed discussion of such formulation to follow-up works.

\subsection{Higher constraints, convergence and approximations}

One important consideration for future works is to develop a more systematic understanding of the convergence of both the structure factor (or more generally the 2RDM) and the energy as we add more constraints and of the kinds of constraints that are best for capturing certain states. We speculate that the $(2,2)$ and $(2,3)$ constraints capture 2-particle and 3-particle correlations to a large extent, leading to accurate ground state properties for states whose main correlations are 2- and 3-body correlations. The prominence of few body correlations is expected since the projected Coulomb Hamiltonian itself is a two-body Hamiltonian. However, there may be more filling-dependent global configurational constraints arising from the structure of the Hilbert space that are not encoded in these constraints, or at least are only encoded on average. One route to understand $(2,p)$ constraints is by viewing them as positivity constraints on 2-body dual Hamiltonians acting on some $p$-particle wavefunctions belonging to a constrained space. While the computational cost for such $(2,p)$ constraints increases with $p$, good approximations of such constraints can be implemented far more efficiently if we can find a good variational ansatz for the space of low energy states of the corresponding Hamiltonian.

 \subsection{Constraints beyond the $(2,p)$ hierarchy}
While we have focused on deriving energetic lower bounds, several generalizations of this approach employed in the context of quantum mechanical bootstrap 
 \cite{HanBootstrappingMatrixQM, Berenstein:2021dyf, PhysRevD.108.125002, berenstein2022bootstrapping,han2020quantum,PhysRevD.107.L051501, cho2022bootstrapping, nancarrow2023bootstrapping}
can also give information about the spectrum and provide bounds on correlation functions. One is to replace the energy minimization by the constraints $\langle \H \rangle = E$ and $\langle \H \mathcal O \rangle = \langle \mathcal O \H \rangle = E \langle \mathcal O \rangle$ for any operator $\O$, which has the effect of restricting to eigenstates of the Hamiltonian. This could potentially yield discrete \emph{windows} of possible values of the energy \cite{nancarrow2023bootstrapping}. For the LLL, the simplest choice for $\O$ involves a product of projected densities $\rho_\bp \rho_{-\bp}$. Such a condition provides a constraint on the 4-RDM rather than the 2RDM. The weaker condition $\langle [\H, \mathcal O] \rangle = 0$ leads to a constraint on the 3-RDM. For more general band systems lacking CMTS, choosing $\O = c^\dagger_\alpha c_\beta$ would lead to constraints on the 3-RDM (using $\langle \H \mathcal O \rangle = \langle \mathcal O \H \rangle = E \langle \mathcal O \rangle$) and 2RDM ($\langle [\H, \mathcal O] \rangle = 0$). Such constraints are normally referred to the \textit{equation of motion (EoM) constraints} since it selects the possible eigenstates of the Hamiltonian~\cite{HanBootstrappingMatrixQM}.

Another possible approach is to notice that the ground state energy cannot be lowered by applying any operator, which implies $\langle \O^\dagger \H \O \rangle \geq E \langle \O^\dagger \O \rangle$ for expectation values evaluated in the ground state. Combined with $\langle \H \O \rangle = E \langle\O \rangle$ which holds for any eigenstate, we get the relation $\langle \O^\dagger [\H, \O] \rangle \geq 0$. If the operator $\O$ can be expanded by a set of basis operators $\{\O_i\}$ such that $\O=\sum_iA_i\O_i$, then the positivity becomes a PSD constraint:
\begin{equation}
    \mathcal{C}\succeq0,\quad\mathcal{C}_{ij}\equiv\langle \mathcal{O}_i^\dagger[\mathcal{H},\mathcal{O}_j]\rangle,
\end{equation}
since we have $ \langle \O^\dagger [\H, \O] \rangle=\sum_{ij}A^*_i\mathcal{C}_{ij}A_j\geq0,\text{ }\forall A$. This PSD constraint is called the \textit{perturbative positivity} that selects (only works for) the ground state (note that the DQG/T1/T2 conditions work for any possible states include the excited states). When considering operators $\O$ for perturbative positivity constraints, it is important to keep track of the number of particles involved since we are studying systems with fixed particle number. For operators which change particle number, such constraints will relate ground state properties at different fillings. We can avoid this by focusing on operators that do not change the particle number. Thus, the choices for $\{\O\}$ are limited to the zero particle basis: order-2 basis $\{ c^\dagger_\alpha c_\beta  \}$ (or $\{\rho_\bq\}$ if the form factor matrix is invertible);  order-4 basis $\{ c^\dagger_\alpha c^\dagger_\beta c_\delta c_\gamma, c^\dagger_\alpha c_\beta\}$; and so on. 
A weaker version of this perturbative positivity is to take the double commutator: $\langle [\O^\dagger, [\H, \O]] \rangle \geq 0$ which will lower the order of RDMs involved by one. For example, if we choose $\{\O\}$ to be $\{\rho_\bq\}$, combining CMTS and GMP algebra, we have the following constraint on the structure factor:
\begin{equation}
    \int\frac{d^2\bp}{2\pi}V_\bp e^{-\frac{p^2}{2}}\sin^2\frac{\bp\wedge\bq}{2}(S_{\bq+\bp}-S_\bp)\geq 0\quad\forall \bq.
\end{equation}

We tested these constraints and found that they do not lead to tighter lower bounds on energy compared to the positivity constraints we use, but they lead to upper bounds on the ground state energy since the perturbative positivity selects the ground state excluding excited states. We found these bounds to be a lot looser than known variational bounds for the simplest operators we tried, but they will likely get tighter as we impose more constraints. Nevertheless, this shows the great potential to get two-side bounds purely from Bootstrap (see also Ref.~\cite{cho2024coarse} for a related discussion). We leave a detailed discussion of these more general bounds to future works. We also note recent attempts aiming to generalize this approach to finite temperatures  \cite{Fawzi2024, cho2024thermal, cho2024coarse}.

\subsection{Higher Landau levels and Chern bands}
A few obvious extensions of this work include (i) higher Landau levels, (ii) bosonic quantum Hall states, and (iii) correlated states in topological flat bands. Direction (i) is the most obvious generalization and can be implemented through the simple replacement $V_\bq \mapsto V_\bq [L_n(\bq^2/2)]^2$. In higher LLs, both non-Abelian states and symmetry broken states such as stripes and Wigner crystals are expected \cite{HaldaneRezayiStripe, HigherLLDMRGPRL, HigherLLDMRG}. Detecting these phases through the structure factor is possible but requires more careful analysis of its dependence as a function of geometry and system size \cite{HigherLLDMRGPRL, HigherLLDMRG}. For direction (ii), we can still employ the Haldane duality and the $(2,2)$ constraints but there is no simple analog of the higher $(2,p)$ constraints. One possibility to address this is to include higher RDMs as variables in the bootstrap approach. For direction (iii), the main difference is the absence of CMTS which means that $\langle \rho_\bq \rho_{-\bq + \bG} \rangle$ is in general non-vanishing for any reciprocal lattice vector $\bG$. As discussed in the main text, we can still derive some version of the Haldane duality under some assumptions that remain to hold for bands which are sufficiently close to the LLL. For bands where the generalized Haldane duality does not hold, we can still perform the bootstrap by using the 2RDM instead of the structure factor. Note that single particle dispersion may also be included, as its expectation value also only depends on the 2RDM.

\section*{Acknowledgements}

We thank Dan Parker for related collaborations. We are grateful to Ashvin Vishwanath, Nick Bultinck, Sid Parameswaran, Debanjan Chowdhury, and Adrian Po for fruitful discussions.
This research was supported in part by grant NSF PHY-2309135 to the Kavli Institute for Theoretical Physics (KITP). This work was performed in part at the Aspen Center for Physics, which is supported by National Science Foundation grant PHY-2210452 and the Simons Foundation (1161654, Troyer). The authors acknowledge supports from FAS Research Computing (FASRC) at Harvard for providing computational resources.

\appendix
\section{Upper bounds from variational Monte Carlo}\label{model_wf}
In this section, we provide the details about the upper bound calculations using variational Monte Carlo (VMC) with model wavefunctions. The method used is called the lattice Monte Carlo for quantum Hall on torus developed by one of us~\cite{Jie_Lattice}. We will not discuss any of the technical details of that method for which the reader is encouraged to read Ref.~\cite{Jie_Lattice}. Here, we outline the model wavefunctions used for the VMC. 

First, for the $\nu=1/m$ Laughlin state, the model wavefunction on torus is given by~\cite{haldane1985periodic}
\begin{equation}
    \Psi_\text{Laughlin}(\{\alpha\}) = \prod_{i<j}^{N}\left[ f(z_i-z_j) \right]^m\prod_{k=1}^mf(Z-\alpha_k),
\end{equation}
where $z=x+iy$ is the complexified electron coordinate, $Z=\sum_i^N z_i$ is the center-of-mass, $\alpha_k$'s $(k=1,\cdots,m)$ are the center-of-mass zeros that encode the periodic boundary condition by requiring $\sum_{k=1}^m\alpha_k = 0\mod \text{BZ}$, and $f(z) = \sigma(z|a_1,a_2)\exp{-\frac{1}{N_\Phi}zz^*}$ with $\sigma(z|a_1,a_2)$ the ``modified Weierstrass sigma function''~\cite{haldane2018modular} ($a_1,a_2$ are two complex vectors defining the unit cell, i.e, the lattice constants).

Then, for the CFL state ($\nu=1/m$ with $m$ even), there are a few model states proposed~\cite{RN413,rezayi2000incompressible,fremling2018trial}. We will use the one in Ref.~\cite{Jie_Lattice} which is more favorable for Monte Carlo calculations.
It is given by~\cite{jain1997quantitative,shao2015entanglement}
\begin{equation}
    \Psi_\text{CFL}(\{\alpha\},\{d\}) = \det \tilde{M}_{ij}\prod_{i<j}\left[ f(z_i-z_j) \right]^{m-2}\prod_{k=1}^mf(Z-\alpha_k),
\end{equation}
where $\tilde{M}_{ij}$ is a $N\times N$ matrix:
\begin{equation}
    \tilde{M}_{ij} = e^{\frac{1}{2m}(z_id_j^*-z_i^*d_j)}\prod_{k\neq j}^N f(z_i-z_k-d_j+\bar{d}),
\end{equation}
with $\det\tilde{M}_{ij}$ its determinant. In addition to a dependence on the $m$ center of mass zeros $\{\alpha\}$,
this wave function depends on $N$ additional parameters $\{d\}$, the dipole moments.

\clearpage
\renewcommand{\theequation}{S\arabic{equation}}
\setcounter{equation}{0}
\renewcommand{\thefigure}{S\arabic{figure}}
\setcounter{figure}{0}
\renewcommand{\thetable}{S\arabic{table}}
\setcounter{table}{0}
\onecolumngrid
\section*{Supplemental Material}
This supplementary material contains detailed discussions on momentum space formulation of the QH problem and the Bootstrapping constraints. 
\vspace{12 pt}
\setcounter{subsection}{0}

\subsection{Bloch states in the LLL and form factors}
We can define the magnetic translation operator $T_\bu = e^{-i \bu \wedge \bR}$. Using the commutation relations of the guiding center $[R_\alpha, R_\beta] = -i \epsilon_{\alpha \beta}$, it is straightforward to show that this operator satisfies the magnetic algebra
\begin{equation}
    T_\bu T_\bv = e^{\frac{i}{2} \bu \wedge \bv} T_{\bu + \bv} = e^{i \bu \wedge \bv} T_{\bv} T_{\bu}
\end{equation}
Here, we used the notation $\wedge \bv = (v_y, -v_x)$ and $\bu \wedge \bv = \bu \cdot (\wedge \bv) = u_x v_y - u_y v_x$. Note that the magnetic translation operator is related to the form factor matrix via $\Lambda_\bq^{\alpha \beta} = \langle \alpha| T_{\wedge \bq}|\beta \rangle$. 

To express the form factor in momentum space, let us define a unit cell containing $2\pi$ flux spanned by the vectors $\ba_1$ and $\ba_2$ such that $|\ba_1 \wedge \ba_2| = 2\pi$. Then for any lattice vectors $\ba = n \ba_1 + m \ba_2$ and $\ba' = n' \ba_1 + m' \ba_2$, the magnetic translation operators commute $[T_\ba, T_{\ba'}] = 0$. As a result, they can be simultaneously diagonalized with the eigenstates corresponding to the Bloch states defined via
\begin{equation}
    T_\ba \psi_\bk = \eta_\ba e^{i \bk \cdot \ba} \psi_\bk
\end{equation}
where the factor $\eta_\ba$, introduced for latter convenience, is $+1$ if $\ba/2$ is a lattice vector and $-1$ otherwise. The explicit action of $T_\bu$ in real space is can be chosen to be $T_\bu f(\br) = e^{-\frac{i}{2} \bu \wedge \br} f(\br + \bu)$. Thus, the Bloch states take the form
\begin{equation}
    \psi_\bk(\br) = C e^{\frac{i}{2} \bar k r} \sigma(r + i k|a_1, a_2) e^{-\frac{1}{4} r \bar r} e^{-\frac{1}{4} k \bar k}
    \label{Bloch}
\end{equation}
where $C$ is an unimportant normalization constant. Here, we use the notation that the complexified version of a variable is denoted by unbolded letter, e.g. $r = r_x + i r_y$, $k = k_x + i k_y$ and $\sigma(z|a_1, a_2)$ is the modified Weierstrass sigma function with lattice parameters $a_1$ and $a_2$ which satisfies
\begin{equation}
    \sigma(z + a|a_1, a_2) = \eta_\ba e^{\frac{\bar a}{2}(z + \frac{a}{2})} \sigma(z|a_1, a_2), \qquad \sigma(-z) = -\sigma(z)
    \label{WeierstrassSigma}
\end{equation}
In what follows, we will drop the lattice parameters and use the simplified notation $\sigma(z) := \sigma(z|a_1, a_2)$.

It is easy to see using (\ref{WeierstrassSigma}) that Bloch states (\ref{Bloch}) satisfy
\begin{equation}
    \psi_\bk(\br + \ba) = \eta_\ba e^{i \bk \cdot \ba} e^{\frac{i}{2} \ba \wedge \br} \phi_\bk(\br), \qquad   \psi_{\bk + \bG}(\br) = \eta_\bG  e^{\frac{i}{2} \bG \wedge \bk} \psi_\bk(\br)
    \label{MagneticBloch}
\end{equation}
In the Bloch basis, the form factor is given by $ \Lambda_\bq^{\bk, \bk'} = \langle \psi_{\bk}|e^{-i \bq \cdot \bR}| \psi_{\bk'} \rangle$. Now notice that $T_\bu e^{i \bq \cdot \bR} = T_\bu T_{-\wedge \bq} = e^{i \bq \cdot \bu} e^{i \bq \cdot \bR} T_\bu$ which means that $T_\ba [e^{i \bq \cdot \bR} \psi_{\bk}] = \eta_\ba e^{i (\bk + \bq) \cdot \ba} e^{i \bq \cdot \bR} \psi_{\bk}$. Thus, since we are considering a single non-degenerate band $e^{i \bq \cdot \bR} \psi_{\bk}$ should be proportional to $\psi_{\bk + \bq}$. The proportionality constant (which generally depends on both $\bk$ and $\bq$) can be determined from the explicit form of the Bloch wavefunctions (\ref{Bloch}) as
\begin{align}
    e^{i \bq \cdot \bR} \psi_\bk(\br) &= T_{-\wedge \bq} \psi_\bk(\br) = C e^{\frac{i}{2} \bq \cdot \br} e^{\frac{i}{2} \bar k (r + iq)} \sigma(r + i (k + q)|a_1, a_2) e^{-\frac{1}{4} (r + iq)(\bar r - i \bar q)} e^{-\frac{1}{4} k \bar k} \nonumber \\
    &= C e^{\frac{i}{2} r (\bar k + \bar q)} \sigma(r + i (k + q)|a_1, a_2) e^{-\frac{1}{4} r \bar r} e^{-\frac{1}{4} k \bar k} e^{-\frac{1}{4} q \bar q - \frac{1}{2} q \bar k} = \psi_{\bk + \bq}(\br) e^{\frac{i}{2} \bq \wedge \bk}, %\quad \implies \quad \lambda_\bq(\bk) = e^{-\frac{i}{2} \bq \wedge \bk}
\end{align}
which leads to the form factor $\Lambda_\bq^{\bk,\bk'} = \langle e^{i \bq \cdot \bR} \psi_\bk|\psi_{\bk'} \rangle = e^{-\frac{i}{2} \bq \wedge \bk} \langle \psi_{\bk + \bq}|\psi_{\bk'} \rangle$. Note that $|\psi_\bk \rangle$ and $|\psi_{\bk'} \rangle$ are not orthogonal unless we restrict $\bk$ and $\bk'$ to the first BZ (if $\bk' = \bk + \bG$, clearly $\langle \psi_{\bk}|\psi_{\bk'} \rangle = \eta_\bG e^{i \bG \wedge \bk} \neq 0$). Let us denote by $[\bk]$ the part of $\bk$ within the first BZ and by $\{\bk\}:= \bk - [\bk]$ its reciprocal lattice part. Using Eq.~\ref{MagneticBloch}, we can write $|\psi_\bk \rangle = \alpha^*_\bk |\psi_{[\bk]} \rangle$ where $\alpha_\bk = \eta_{\{\bk\}} e^{\frac{i}{2} [\bk] \wedge \{\bk\}}$ which leads to the expression of the form factor
\begin{equation}
    \Lambda_\bq^{\bk,\bk'} = e^{-\frac{i}{2} \bq \wedge \bk} \langle \psi_{\bk + \bq}|\psi_{\bk'} \rangle = e^{-\frac{i}{2} \bq \wedge \bk} \alpha^*_{\bk'} \alpha_{\bk + \bq} \delta_{[\bk'],[\bk + \bq]} %= \lambda_\bq(\bk) \delta_{[\bk'],[\bk + \bq]}, \qquad \lambda_\bq(\bk) = e^{-\frac{i}{2} \bq \wedge \bk} \alpha^*_{\bk'} \alpha_{\bk + \bq}
\end{equation}
Notice that the form factor is not gauge invariant. Instead, it changes by $\Lambda_\bq^{\bk,\bk'} \mapsto \Lambda_\bq^{\bk,\bk'} e^{i [\phi(\bk') - \phi(\bk)]}$ under the gauge transformation $\psi_\bk \mapsto e^{i \phi(\bk)} \psi_\bk$. We will find it convenient to choose a periodic gauge corresponding to the wavefunctions
\begin{equation}
    |\Tilde{\psi}_{\bk} \rangle \equiv \alpha_{\bk}|\psi_\bk \rangle = |\psi_{[\bk]} \rangle
\end{equation}
which is manifestly periodic in $\bk$ since it depends only on $[\bk]$. The form factors in the periodic gauge become
\begin{equation}
    \Lambda_\bq^{\bk,\bk'} = e^{-\frac{i}{2} \bq \wedge \bk} \alpha_\bk^* \alpha_{\bk'} \langle \psi_{\bk + \bq}|\psi_{\bk'} \rangle = e^{-\frac{i}{2} \bq \wedge \bk} \alpha^*_{\bk} \alpha_{\bk + \bq} \delta_{[\bk'],[\bk + \bq]} = \lambda_\bq(\bk) \delta_{[\bk'],[\bk + \bq]}, \qquad \lambda_\bq(\bk) = e^{-\frac{i}{2} \bq \wedge \bk} \alpha^*_{\bk} \alpha_{\bk + \bq}
\end{equation}
This expression is manifestly invariant under $\bk' \mapsto \bk' + \bG$ for any reciprocal lattice vector $\bG$. To show it is also periodic in $\bk$, note that for any reciprocal lattice vectors $e^{\pm \frac{i}{2} \bG \wedge \bG'} = \eta_\bG \eta_{\bG'} \eta_{\bG + \bG'}$ which implies
\begin{equation}
    \alpha_{\bk + \bG} = \eta_{\{\bk\}} \eta_{\{\bk\} + \bG} e^{\frac{i}{2} [\bk] \wedge \bG} \alpha_{\bk} = \eta_{\bG} e^{\frac{i}{2} \bk \wedge \bG} \alpha_{\bk}, 
\end{equation}
leading to the identities
\begin{equation}
    \lambda_\bq(\bk + \bG) = \lambda_\bq(\bk)\qquad \lambda_{\bq + \bG}(\bk) = \eta_\bG e^{\frac{i}{2}\bq \wedge \bG} e^{-i \bG \wedge \bk} \lambda_\bq(\bk).
    \label{lambdaIdentities}
\end{equation}

\subsection{Continuous magnetic translation symmetry (CMTS)}
The action of magnetic translation symmetry in the periodic gauge is
\begin{equation}
    T_{-\wedge\bq} \Tilde{\psi}_{\bk}(\br) = \alpha_\bk T_{-\wedge\bq} \psi_\bk(\br) = \alpha_\bk\psi_{\bk+\bq}(\br)e^{-\frac{i}{2}\bk\wedge\bq}\\
     = \alpha_\bk\alpha^*_{\bk+\bq}e^{-\frac{i}{2}\bk\wedge\bq}\Tilde{\psi}_{\bk+\bq}(\br)
\end{equation}

by noting that $\alpha_{[\bk]} = 1$.

To see how the second-quantized operator transforms under the magnetic translation, we first look at the transformation of a single fermion operator in position space:
\begin{equation}
    \mathcal{T}_\bu^{-1} c_\br \mathcal{T}_\bu = e^{-\frac{i}{2}\bu\wedge\br} c_{\br+\bu} = T_\bu c_\br.
\end{equation}
We can then expand $c_\br$ using the momentum basis:
\begin{equation}
    c_\br = \sum_{\bk\in \text{BZ}}\Tilde{\psi}_{\bk}(\br)c_\bk \quad \text{ with } \quad c_{\bk+\bG} = c_\bk.
\end{equation}

Thus, the magnetic translation becomes
\begin{align}
    \sum_{\bk\in \text{BZ}}\Tilde{\psi}_{\bk}(\br)\mathcal{T}_{-\wedge\bq}^{-1}c_\bk\mathcal{T}_{-\wedge\bq} &= \sum_{\bk\in \text{BZ}}\left[ T_{-\wedge\bq} \Tilde{\psi}_{\bk}(\br) \right] c_\bk = \sum_{\bk\in \text{BZ}}\alpha_\bk\alpha^*_{\bk+\bq}e^{-\frac{i}{2}\bk\wedge\bq} \Tilde{\psi}_{\bk+\bq}(\br) c_\bk \nonumber \\
    & = \sum_{\bk\in \text{BZ}} \lambda^*_\bq(\bk) \Tilde{\psi}_{\bk+\bq}(\br) c_\bk = \sum_{\bk\in \text{BZ}}\lambda^*_\bq(\bk - \bq) \Tilde{\psi}_{\bk}(\br) c_{\bk-\bq} = \sum_{\bk\in \text{BZ}}\lambda_{-\bq}(\bk) \Tilde{\psi}_{\bk}(\br) c_{\bk-\bq}
\end{align}
which implies 
\begin{equation}
    \mathcal{T}_{-\wedge\bq}^{-1}c_\bk\mathcal{T}_{-\wedge\bq} = \lambda_{-\bq}(\bk) c_{\bk-\bq} \equiv \beta_{\bk,\bq}c_{\bk-\bq}.
\end{equation}
Similarly, we have
\begin{equation}
    \mathcal{T}_{-\wedge\bq}^{-1}c^\dagger_\bk\mathcal{T}_{-\wedge\bq} =\beta^*_{\bk,\bq}c^\dagger_{\bk-\bq}.
\end{equation}

Now, let us discuss the continuous magnetic translation symmetry (CMTS) for the density matrices.
First we define the density operator in the following way:
\begin{equation}
    \hat{\mathcal{D}}^{n} = d^\dagger_{\sigma_1,\bk_1}\cdots d^\dagger_{\sigma_n,\bk_n} d_{\sigma_n,\bk'_n}\cdots d_{\sigma_1,\bk'_1},
\end{equation}
where $\sigma_i = \pm$ for $i=1,\cdots,n$ and $d_{+,\bk} = c_\bk$ and $d_{-,\bk} = c^\dagger_{\bk}$. Thus, the CMTS in terms of $d$'s becomes
\begin{equation}
    \mathcal{T}_{-\wedge\bq}^{-1}d_{\sigma,\bk}\mathcal{T}_{-\wedge\bq} =\beta^{\sigma}_{\bk,\bq}d_{\sigma,\bk-\bq}\quad\text{ with }\quad \beta^+ = \beta \text{ and } \beta^- = \beta^*.
\end{equation}
We can call this density operator the order $(n-m,m)$ density operator and denote it as $\hat{\mathcal{D}}^{(n,m)}$, where $m = \sum_{l=1}^n(|\sigma_l|+\sigma_l)/2$.
It is not hard to see that the density operator is not invariant under the continuous magnetic translation. We have to look at the physical state that is invariant under such transformation. Given a state $|\Psi\rangle$ such that
\begin{equation}
    \mathcal{T}_\bu |\Psi\rangle = e^{i\theta_\bu}|\Psi\rangle,
\end{equation}
we should have
\begin{equation}
    \langle\Psi|\mathcal{T}_\bu^{-1} \hat{\mathcal{D}}^{(n,m)} \mathcal{T}_\bu |\Psi\rangle = \langle\Psi|e^{-i\theta_\bu} \hat{\mathcal{D}}^{(n,m)} e^{i\theta_\bu} |\Psi\rangle = \langle\Psi| \hat{\mathcal{D}}^{(n,m)}  |\Psi\rangle ,
\end{equation}
which gives the CMTS in the primal form.

Now, if the ground state $|\Omega\rangle$ of the system doesn't spontaneously break the CMTS~\cite{Note14}, we can define the density matrix in momentum space corresponding to $\hat{\mathcal{D}}^{(n,m)} $ as
\begin{equation}
    \mathcal{M}^{(n,m)}_{\bk_1,\cdots,\bk_n;\bk'_1,\cdots,\bk'_n}\equiv \langle\Omega| d^\dagger_{\sigma_1,\bk_1}\cdots d^\dagger_{\sigma_n,\bk_n} d_{\sigma_n,\bk'_n}\cdots d_{\sigma_1,\bk'_1}|\Omega\rangle.
\end{equation}
The CMTS then implies that
\begin{equation}\label{CMTS}
    \mathcal{M}^{(n,m)}_{\bk_1,\cdots,\bk_n;\bk'_1,\cdots,\bk'_n} = \left[\prod_{l=1}^n\left(\beta^{\sigma_l}_{\bk_l,\bq}\right)^*\beta^{\sigma_l}_{\bk'_l,\bq}\right]\mathcal{M}^{(n,m)}_{\bk_1-\bq,\cdots,\bk_n-\bq;\bk'_1-\bq,\cdots,\bk'_n-\bq}
\end{equation}
If we further consider the crystal momentum conservation, we can define the total momentum $\bQ$ as
\begin{equation}
    \bQ = \left[\sum_{l=1}^n \sigma_l\bk_l\right] = \left[\sum_{l=1}^n \sigma_l\bk'_l\right],
\end{equation}
thus we can eliminate two momentum variables from the two equalities above:
\begin{equation}
    \bk_n = \sigma_n\left[ \bQ-\sum_{l=1}^{n-1}\sigma_l\bk_l \right]\quad\text{ and }\quad \bk'_n = \sigma_n\left[ \bQ-\sum_{l=1}^{n-1}\sigma_l\bk'_l \right].
\end{equation}
Substitute into Eq.~\eqref{CMTS}, we have the transformation between different momentum sectors:
\begin{equation}
\begin{split}
   & \mathcal{M}^{(n,m),\bQ}_{\bk_1,\cdots,\bk_{n-1};\bk'_1,\cdots,\bk'_{n-1}}\\
   &= \left(\beta^{\sigma_n}_{\sigma_n[ \bQ-\sum_{l=1}^{n-1}\sigma_l\bk_l ],\bq}\right)^*\beta^{\sigma_n}_{\sigma_n[ \bQ-\sum_{l=1}^{n-1}\sigma_l\bk'_l ],\bq}\left[\prod_{l=1}^{n-1}\left(\beta^{\sigma_l}_{\bk_l,\bq}\right)^*\beta^{\sigma_l}_{\bk'_l,\bq}\right]\mathcal{M}^{(n,m),[\bQ-(2m-n)\bq]}_{\bk_1-\bq,\cdots,\bk_{n-1}-\bq;\bk'_1-\bq,\cdots,\bk'_{n-1}-\bq}.
\end{split}
\end{equation}
Although the mapping looks complicated, it simply states that the two matrices are related in the following way:
\begin{equation}
    \mathcal{M}^{(n,m),\bQ}(\bk,\bk') = \mathcal{U}^{\bQ\dagger}(\bk,\bk') \mathcal{M}^{(n,m),[\bQ-(2m-n)\bq]}(\bk-\bq,\bk'-\bq)\mathcal{U}^\bQ(\bk,\bk'),
\end{equation}
where $ \bk=\{\bk_1,\bk_2,\cdots,\bk_{n-1}\}$ and $\mathcal{U}^\bQ$ is a diagonal unitary matrix in the indices $(\bk,\bk')$. Notice that both the constant index shifting and the unitary transformation cannot change the spectrum of a matrix. Thus, under the CMTS, $\mathcal{M}^{(n,m),\bQ} $ and $ \mathcal{M}^{(n,m),[\bQ-(2m-n)\bq]}$ have exactly the same spectrum.

\subsection{Momentum space formulation of the $(2,p)$ constraints}
In this section, we show in detail how to write the positivity constraints on a momentum basis.
\subsubsection{Inverse form factor matrix for 0$^\text{th}$LL and (2,2) positivity constraints}
Given that in the momentum space
\begin{equation}
    \rho_\bq = \sum_\bk c^\dagger_\bk c_{\bk+\bq}\lambda_\bq(\bk) = e^{-\frac{1}{4}\bq^2}\sum_\bk c^\dagger_\bk c_{\bk+\bq}\tilde\lambda_\bq(\bk)\equiv e^{-\frac{1}{4}\bq^2}\tilde\rho_\bq,
\end{equation}
or equivalently,
\begin{equation}\label{form_factor_matrix}
    \tilde\rho_\bq = \sum_{\bk,\bk'}\Lambda_\bq^{\bk,\bk'}c^\dagger_\bk c_{\bk'}\quad\text{with} \quad\Lambda_\bq^{\bk,\bk'}\equiv\delta_{[\bq+\bk-\bk'],0} \tilde\lambda_\bq(\bk),
\end{equation}
we shall have
\begin{equation}
    \tilde\rho_{\bq+\bG} = \sum_\bk c^\dagger_\bk c_{\bk+\bq}\tilde\lambda_{\bq+\bG}(\bk) = \sum_\bk c^\dagger_\bk c_{\bk+\bq}\eta_\bG e^{-i\bG\wedge \bk+\frac{i}{2}\bq\wedge\bG}\tilde\lambda_\bq(\bk),
\end{equation}
where we used
\begin{equation}
    \begin{split}
        \tilde\lambda_{\bq+\bG}(\bk) &= e^{-\frac{i}{2}(\bq+\bG)\wedge\bk}\alpha^*_\bk\alpha_{\bk+\bq+\bG} = e^{-\frac{i}{2}(\bq+\bG)\wedge\bk}\eta_\bG e^{\frac{i}{2}(\bk+\bq)\wedge\bG}\alpha^*_\bk\alpha_{\bk+\bq} \\
        & = \eta_\bG e^{-i\bG\wedge \bk+\frac{i}{2}\bq\wedge\bG}\tilde\lambda_\bq(\bk).
    \end{split}
\end{equation}
This then leads to
\begin{equation}
    \tilde\rho_{\bq+\bG}\eta_\bG e^{-\frac{i}{2}\bq\wedge\bG} = \sum_\bk c^\dagger_\bk c_{\bk+\bq} e^{-i\bG\wedge \bk}\tilde\lambda_\bq(\bk).
\end{equation}
Multiplying both sides by $e^{i\bG\wedge \bk'}\tilde\lambda_\bq^*(\bk')/N_\bG$ ($N_\bG = N_\Phi$ in this case) and summing over $\bG$, we get
\begin{equation}
    \begin{split}
        \frac{1}{N_\bG} \sum_\bG e^{i\bG\wedge \bk'}\tilde\rho_{\bq+\bG}\eta_\bG e^{-\frac{i}{2}\bq\wedge\bG}\tilde\lambda_\bq^*(\bk')
        =& \sum_\bk c^\dagger_\bk c_{\bk+\bq} \tilde\lambda_\bq(\bk)\tilde\lambda_\bq^*(\bk')\frac{1}{N_\bG} \sum_\bG e^{i\bG\wedge (\bk'-\bk)} \\
        =& \sum_\bk c^\dagger_\bk c_{\bk+\bq} \tilde\lambda_\bq(\bk)\lambda_\bq^*(\bk')\delta_{\bk,\bk'}\\
        =& c^\dagger_{\bk'} c_{\bk'+\bq},
    \end{split}
\end{equation}
where we have used the fact that $ \frac{1}{N_\bG} \sum_\bG e^{i\bG\wedge (\bk'-\bk)} = \delta_{\bk,\bk'}$ and $ | \tilde\lambda_\bq(\bk)| = 1$ by construction.
Notice that $c^\dagger_\bk c_{\bk+\bq} \equiv c^\dagger_\bk c_{\bk+[\bq]}$ on the right-hand side, we can restrict $\bq$ to be within the first BZ and let $\bq'\equiv\bq+\bG$ to be the unrestricted momentum (thus, $ \bq = [\bq]$ and $\bG = \{\bq'\}$). By doing this, the above equation can be rewritten in the following form:

\begin{equation}
    \begin{split}
        c^\dagger_{\bk} c_{\bk+\bq} = \frac{1}{N_\bG}\sum_{\bq'} \delta_{[\bq'-\bq],0}e^{i\{\bq'\}\wedge \bk} \tilde\rho_{\bq'}\eta_{\{\bq'\}} e^{-\frac{i}{2}[\bq']\wedge\{\bq'\}}\lambda_{[\bq']}^*(\bk).
    \end{split}
\end{equation}
Then, let $\bk'=\bk+\bq$, we can further adopt a matrix form for such expression:
\begin{equation}
    c^\dagger_{\bk} c_{\bk'} = \sum_\bq\Lambda^\bq_{\bk,\bk'}\tilde\rho_\bq
\end{equation}
with the matrix $ \Lambda^\bq_{\bk,\bk'}$ defined as
\begin{equation}\label{form_factor_matrix_inv}
\begin{split}
    \Lambda^\bq_{\bk,\bk'}\equiv&\frac{1}{N_\bG}\delta_{[\bq-(\bk'-\bk)],0}e^{i\{\bq\}\wedge \bk}e^{-\frac{i}{2}[\bq]\wedge\{\bq\}}\eta_{\{\bq\}}\tilde\lambda_{[\bq]}^*(\bk) \\
    =& \frac{1}{N_\bG}\delta_{[\bq+\bk-\bk'],0}\tilde\lambda_{\bq}^*(\bk).
\end{split}
\end{equation}
One can check that 
\begin{equation}\label{Orthogonality1}
\begin{split}
    \sum_\bq \Lambda^\bq_{\bk_1,\bk_2}\Lambda_\bq^{\bk_3,\bk_4} =& \frac{1}{N_\bG}\sum_{\bq} \delta_{[\bq+\bk_1-\bk_2],0}\delta_{[\bq+\bk_3-\bk_4],0} \tilde\lambda^*_\bq(\bk_1)\tilde\lambda_\bq(\bk_3)  \\
    =& \sum_{[\bq]} \delta_{[[\bq]+\bk_1-\bk_2],0}\delta_{[[\bq]+\bk_3-\bk_4],0} \frac{1}{N_\bG}\sum_\bG \tilde\lambda^*_{[\bq]+\bG}(\bk_1)\tilde\lambda_{[\bq]+\bG}(\bk_3) \\
    =& \sum_{[\bq]} \delta_{[[\bq]+\bk_1-\bk_2],0}\delta_{[[\bq]+\bk_3-\bk_4],0}  \tilde\lambda^*_{[\bq]}(\bk_1)\tilde\lambda_{[\bq]}(\bk_3)\frac{1}{N_\bG}\sum_\bG e^{i\bG\wedge(\bk_1-\bk_3)} \\
    =& \sum_{[\bq]} \delta_{[[\bq]+\bk_1-\bk_2],0}\delta_{[[\bq]+\bk_3-\bk_4],0} \delta_{\bk_1,\bk_3} = \delta_{\bk_1,\bk_3}\delta_{\bk_2,\bk_4}.
\end{split}
\end{equation}
and
\begin{equation}\label{Orthogonality2}
\begin{split}
    \sum_{\bk_1,\bk_2} \Lambda^\bq_{\bk_1,\bk_2}\Lambda_{\bq'}^{\bk_1,\bk_2} =& \frac{1}{N_\bG}\sum_{\bk_1,\bk_2} \delta_{[\bq+\bk_1-\bk_2],0}\delta_{[\bq'+\bk_1-\bk_2],0} \tilde\lambda^*_\bq(\bk_1)\tilde\lambda_{\bq'}(\bk_1)  \\
    =& \frac{1}{N_\bG}\sum_{\bk_1} \delta_{[\bq-\bq'],0}\tilde\lambda^*_\bq(\bk_1)\tilde\lambda_{\bq'}(\bk_1) \\
    =& \frac{1}{N_\bG}\sum_{\bk_1} \delta_{[\bq-\bq'],0}\sum_\bG\tilde\lambda^*_\bq(\bk_1)\tilde\lambda_{\bq+\bG}(\bk_1) \delta_{\bG,\{\bq'-\bq\}}\\
    =&  \delta_{[\bq-\bq'],0}\sum_\bG\delta_{\bG,\{\bq'-\bq\}}\frac{1}{N_\bG}\sum_{\bk_1}\eta_\bG e^{i\bG\wedge \bk_1-\frac{i}{2}\bq\wedge\bG} \\
    =& \delta_{[\bq-\bq'],0}\delta_{\{\bq-\bq'\},0} = \delta_{\bq,\bq'}.
\end{split}
\end{equation}

The expectation value of the two-particle operator $c^\dagger_{\bk_1}c_{\bk_2}c^\dagger_{\bk_3}c_{\bk_4}$ can then be related to the structure factor in the following way:
\begin{equation}
\begin{split}
    \langle c^\dagger_{\bk_1}c_{\bk_2}c^\dagger_{\bk_3}c_{\bk_4}\rangle =& \sum_{\bq,\bp}\Lambda^\bq_{\bk_1,\bk_2}\Lambda^\bp_{\bk_3,\bk_4}\langle \tilde\rho_\bq\tilde\rho_\bp\rangle =  \sum_{\bq,\bp}\Lambda^\bq_{\bk_1,\bk_2}\Lambda^{-\bq}_{\bk_3,\bk_4}\langle \tilde\rho_\bq\tilde\rho_{-\bq}\rangle\delta_{\bp,-\bq}\\
    =& \sum_{\bq}\Lambda^\bq_{\bk_1,\bk_2}\Lambda^{-\bq}_{\bk_3,\bk_4}S_\bq \\
    =& \frac{1}{N_G^2}\sum_\bq \delta_{[\bq+\bk_1-\bk_2],0}\delta_{[-\bq+\bk_3-\bk_4],0}\lambda_{\bq}^*(\bk_1)\lambda_{-\bq}^*(\bk_3)S_\bq \\
    =& \frac{1}{N_G^2}\sum_\bq \delta_{[\bq+\bk_1-\bk_2],0}\delta_{[-\bq+\bk_3-\bk_4],0}\tilde\lambda_{\bq}^*(\bk_1)\tilde\lambda^*_{-\bq}(\bk_3)S_\bq .
\end{split}
\end{equation}
Let us now construct matrices that have indices $\bq$ and $\bk\bk'$:
\begin{equation}
    [M_{12}]_{\bk_1\bk_2,\bq} \equiv \frac{1}{N_G} \delta_{[\bq+\bk_1-\bk_2],0}\tilde\lambda_{\bq}^*(\bk_1) \quad \text{and} \quad [M_{34}]_{\bq,\bk_3\bk_4} \equiv \frac{1}{N_G} \delta_{[-\bq+\bk_3-\bk_4],0}\tilde\lambda_{-\bq}^*(\bk_3)
\end{equation}
since there are $N_G^2$ numbers of $\bq$ points and $N_G$ numbers of $\bk$ points so that the matrices constructed are square. Using the same reasoning in Eqs.~\eqref{Orthogonality1},\eqref{Orthogonality2}, we find $M_{12} = M_{34}^\dagger $ and
\begin{equation}
\begin{split}
    \left[M_{12}^\dagger M_{12}\right]_{\bq,\bq'} = \frac{1}{N_G}\delta_{\bq,\bq'}, \qquad
    \left[M_{12} M_{12}^\dagger\right]_{\bk_1\bk_2,\bk_3\bk_4} = \frac{1}{N_G}\delta_{\bk_2\bk_1,\bk_3\bk_4},
\end{split}
\end{equation}
where $\delta_{\bk_1\bk_2,\bk_3\bk_4}\equiv \delta_{\bk_1,\bk_3}\delta_{\bk_2,\bk_4} $. This follows from the fact that $\left[ \delta_{[-\bq+\bk_3-\bk_4],0}\tilde\lambda_{-\bq}^*(\bk_3) \right]^* = \delta_{[-\bq+\bk_3-\bk_4],0}\tilde\lambda_{-\bq}(\bk_3) = \delta_{[-\bq+\bk_3-\bk_4],0}\tilde\lambda^*_{\bq}(\bk_3-\bq) = \delta_{[\bq+\bk_4-\bk_3],0}\tilde\lambda^*_{\bq}(\bk_4)$, thus $\left[[M_{34}]_{\bq,\bk_3\bk_4}\right]^* = [M_{12}]_{\bk_4\bk_3,\bq}\Rightarrow M_{34}^\dagger = M_{12}$. Be noted that the $\bk$ labels are also flipped.
Thus, $M_{12}$ $(M_{34})$ is actually a unitary matrix up to a constant scaling. In this sense, the two-particle expectation value is given by
\begin{equation}
    \langle c^\dagger_{\bk_1}c_{\bk_2}c^\dagger_{\bk_3}c_{\bk_4}\rangle = \left[ \sqrt{N_\bG}M_{12}\diag[ S_\bq/N_\bG  ] \sqrt{N_\bG}M_{12}^\dagger\right]_{\bk_2\bk_1,\bk_3\bk_4},
\end{equation}
which is a unitary rotation of matrix $\diag[S_\bq /N_\bG ]$. Thus, $S_\bq\geq 0$ for all $\bq$ implies that the two-particle expectation value in the form of a matrix with indices $\bk_1\bk_2$ and $\bk_4\bk_3$ is necessarily positive semi-definite (PSD). This can be summarized as
\begin{equation}
    \langle c^\dagger c c^\dagger c \rangle\in \mathcal{H}^{N_\bq}_{+} \Leftrightarrow S_\bq\geq 0\quad\forall \bq,
\end{equation}
where $\mathcal{H}^{N_q}_{+} $ is the set of PSD hermitian matrices with dimension $N_\bq\times N_\bq$.

However, for other types of four-fermion operators like $\langle ccc^\dagger c^\dagger \rangle$ and $\langle c^\dagger c^\dagger cc \rangle$, this is in general not true.
For example, let us consider the four-fermion operator $\langle c^\dagger c^\dagger cc \rangle$. Notice that because of the fermionic anti-commutation relation, we should restrict the indices. We can define
\begin{equation}
    \mathcal{D}^{\bk_1<\bk_2}_{\bk_3<\bk_4} \equiv \langle c^\dagger_{\bk_2}c^\dagger_{\bk_1}c_{\bk_3}c_{\bk_4}\rangle,
\end{equation}
where $\bk_1<\bk_2$ just denotes a way of pairing $\bk_1$ and $\bk_2$ to avoid its reverse and bears no meaning of the magnitude comparison. By definition, the matrix $\mathcal{D} $ has a dimension $ \frac{N_\bk(N_\bk-1)}{2}\times \frac{N_\bk(N_\bk-1)}{2}$ (note: $N_\bq = N_\bk^2=N_\bG^2=N_\Phi^2$). Then, for a two-hole operator $\O_1 = \sum_{\bk_1<\bk_2}A_{\bk_1\bk_2}c_{\bk_1}c_{\bk_2}$, we have
\begin{equation}
    \langle \O_1^\dagger\O_1 \rangle = \sum_{\bk_1<\bk_2,\bk_3<\bk_4}A^*_{\bk_1\bk_2}\mathcal{D}^{\bk_1<\bk_2}_{\bk_3<\bk_4}A_{\bk_3\bk_4} \geq 0\quad\forall A\in\mathbb{C}^{N_\bk(N_\bk-1)/2},
\end{equation}
which means $\mathcal{D}\in\mathcal{H}^{N_\bk(N_\bk-1)/2}_{+}$. Using the expressions derived above, we can rewrite the matrix $\mathcal{D}$ as
\begin{equation}
    \mathcal{D}^{\bk_1<\bk_2}_{\bk_3<\bk_4} = -\langle c^\dagger_{\bk_2}c^\dagger_{\bk_1}c_{\bk_4}c_{\bk_3}\rangle = \langle c^\dagger_{\bk_2}c_{\bk_4}c^\dagger_{\bk_1}c_{\bk_3}\rangle - \nu \delta_{\bk_2,\bk_3}\delta_{\bk_1,\bk_4} = \sum_{\bq}\Lambda^\bq_{\bk_2,\bk_4}\Lambda^{-\bq}_{\bk_1,\bk_3}S_\bq- \nu \delta_{\bk_2,\bk_3}\delta_{\bk_1,\bk_4}.
\end{equation}
In this case, $S_\bq \geq 0$ implies nothing about the semi-definiteness of the matrix $\mathcal{D}$ which therefore acts non-trivially on the Bootstrapping process.

The other four-fermion operator $\langle ccc^\dagger c^\dagger \rangle$ is actually related to the above discussed one by the particle-hole symmetry and gives weaker constraints when $\nu< 1/2$ and exactly the same constraints when $\nu=1/2$. Let us define
\begin{equation}
    \Q^{\bk_1<\bk_2}_{\bk_3<\bk_4} \equiv \langle c_{\bk_3}c_{\bk_4}c^\dagger_{\bk_2}c^\dagger_{\bk_1}\rangle,
\end{equation}
which can be evaluated as
\begin{equation}
    \Q^{\bk_1<\bk_2}_{\bk_3<\bk_4} = \langle c^\dagger_{\bk_2}c_{\bk_4}c^\dagger_{\bk_1}c_{\bk_3}\rangle + (1-2\nu) \delta_{\bk_1,\bk_3}\delta_{\bk_2,\bk_4} - (1-\nu)\delta_{\bk_2,\bk_3}\delta_{\bk_1,\bk_4}.
\end{equation}
One can see imediately that $ \Q^{\bk_1<\bk_2}_{\bk_3<\bk_4}(\nu=1/2) = \mathcal{D}^{\bk_1<\bk_2}_{\bk_3<\bk_4}(\nu=1/2)$. To see this only gives weaker constraints for $\nu<1/2$, we note that by the way we vectorized the coefficients $A_{\bk_1,\bk_2} $ (or anti-symmetrization), $\delta_{\bk_2,\bk_3}\delta_{\bk_1,\bk_4}$ is always zero since $ \bk_2=\bk_3$ and $\bk_1=\bk_4$ are incompatible with the conditions $ \bk_1<\bk_2$ and $\bk_3<\bk_4 $. Moreover, $ \delta_{\bk_1,\bk_3}\delta_{\bk_2,\bk_4}$ is the identity matrix in our notation. So. all together, we have simply:
\begin{equation}
    \Q^{\bk_1<\bk_2}_{\bk_3<\bk_4} = \mathcal{D}^{\bk_1<\bk_2}_{\bk_3<\bk_4} + (1-2\nu)\mathcal{I}.
\end{equation}
This means that, for $\nu<1/2$, a positive semi-definite $\mathcal{D}^{\bk_1<\bk_2}_{\bk_3<\bk_4}$ always leads to a positive definite $\Q^{\bk_1<\bk_2}_{\bk_3<\bk_4} $.

\subsubsection{GMP algebra in terms of inverse form factor matrix}
The GMP algebra is also encoded in the inverse form factor matrix but in a much more complicated way. To see this, we first observe that
\begin{equation}
    c^\dagger_{\bk_1}c_{\bk_2}c^\dagger_{\bk_3}c_{\bk_4} - c^\dagger_{\bk_3}c_{\bk_4}c^\dagger_{\bk_1}c_{\bk_2} = c^\dagger_{\bk_1}c_{\bk_4}\delta_{\bk_2,\bk_3} - c^\dagger_{\bk_3}c_{\bk_2}\delta_{\bk_1,\bk_4},
\end{equation}
which then can be rewritten in terms of inverse form factor matrices:
\begin{equation}
    \sum_{\bq,\bp}\Lambda^\bq_{\bk_1,\bk_2}\Lambda^\bp_{\bk_3,\bk_4}\left[\tilde\rho_\bq,\tilde\rho_\bp\right]  =\sum_{\bq,\bp}\Lambda^\bq_{\bk_1,\bk_2}\Lambda^\bp_{\bk_3,\bk_4}\tilde\rho_{\bq+\bp}2i\sin\frac{\bq\wedge\bp}{2} =  \sum_\bq \left( \Lambda^\bq_{\bk_1,\bk_4}\delta_{\bk_2,\bk_3}-\Lambda^\bq_{\bk_3,\bk_2}\delta_{\bk_1,\bk_4} \right)\tilde\rho_\bq,
\end{equation}
which can be rewritten as
\begin{equation}
    \sum_\bq\left(\sum_\bp \Lambda^{\bq-\bp}_{\bk_1,\bk_2}\Lambda^\bp_{\bk_3,\bk_4}2i\sin\frac{\bq\wedge\bp}{2} \right)\tilde\rho_{\bq}=\sum_\bq \left( \Lambda^\bq_{\bk_1,\bk_4}\delta_{\bk_2,\bk_3}-\Lambda^\bq_{\bk_3,\bk_2}\delta_{\bk_1,\bk_4} \right)\tilde\rho_\bq,\quad\forall \bk_{1,2,3,4}\in\text{BZ}.
\end{equation}
Multiplying both sides by $\tilde\rho_{-\bq'}$ and taking the expectation value, we have
\begin{equation}\label{GMP_equality_constraint}
    \left(\sum_\bp \Lambda^{\bq-\bp}_{\bk_1,\bk_2}\Lambda^\bp_{\bk_3,\bk_4}2i\sin\frac{\bq\wedge\bp}{2} \right)S_{\bq}=\left( \Lambda^\bq_{\bk_1,\bk_4}\delta_{\bk_2,\bk_3}-\Lambda^\bq_{\bk_3,\bk_2}\delta_{\bk_1,\bk_4} \right)S_\bq,\quad\forall \bk_{1,2,3,4}\in\text{BZ}.
\end{equation}
Notice that we have in general $S_\bq>0$, the GMP algebra becomes simply:
\begin{equation}\label{GMP_equality_constraint}
    \sum_\bp \Lambda^{\bq-\bp}_{\bk_1,\bk_2}\Lambda^\bp_{\bk_3,\bk_4}2i\sin\frac{\bq\wedge\bp}{2}= \Lambda^\bq_{\bk_1,\bk_4}\delta_{\bk_2,\bk_3}-\Lambda^\bq_{\bk_3,\bk_2}\delta_{\bk_1,\bk_4} ,\quad\forall \bk_{1,2,3,4}\in\text{BZ}.
\end{equation}
One should be aware that this is not a constraint but a property of the inverse form factor matrix for the 0$^\text{th}$LL, since no $S_\bq$ dependence is in it. This also means by using the inverse form factor matrix, the GMP algebra for the $S_\bq$ (or 2RDM) is built in, and does not need to be implemented as a separate constraint.

Another way of seeing this is the following. Consider an arbitrary bilinear operator
\begin{equation}
    \hat A = \sum_{\alpha \beta} A_{\alpha \beta} c_\alpha^\dagger c_\beta
\end{equation}
where $A$ denotes some operator, $\hat A$ is its second quantized representation and $A_{\alpha \beta}$ are its matrix elements in some basis. It is straightforward to show that $[\hat A, \hat B] = \widehat{[A,B]}$ and that $\hat A = \hat B$ if and only if $A = B$ (which also means $A_{\alpha \beta} = B_{\alpha \beta}$ for all $\alpha$ and $\beta$). Furthermore, the form factors are the matrix representation of the density operator $\tilde \rho_\bq = \hat \Lambda_\bq$. Thus, the GMP algebra implies
\begin{equation}
    \widehat{[\Lambda_\bq,\Lambda_\bp]} = 2i \sin \frac{\bq \wedge \bp}{2} \hat \Lambda_{\bp + \bq} \quad \leftrightarrow \quad [\Lambda_\bq,\Lambda_\bp] = 2i \sin \frac{\bq \wedge \bp}{2} \Lambda_{\bp + \bq}
\end{equation}
where the last relation holds as an identity for the matrix elements. Thus, the GMP algebra is encoded in the form factor matrix. Then, one can use the relation between the form factor matrix and its inverse to get the expression~\eqref{GMP_equality_constraint}.

\subsubsection{(2,3)-T1/T2 constraint}
In this section, we want to impose the so-called (2,3)-T1 constraint which is deduced from convexly combining two three-particle positivity constraints to get a reduced constraint for two-particle expectation values. Consider the following two operators: $\O_{T1,1} = \sum_{\bk_1,\bk_2,\bk_3}A_{\bk_1\bk_2\bk_3}c_{\bk_1}c_{\bk_2}c_{\bk_3}$ and $\O_{T1,2} = \sum_{\bk_1,\bk_2,\bk_3}A^*_{\bk_1\bk_2\bk_3}c^\dagger_{\bk_1}c^\dagger_{\bk_2}c^\dagger_{\bk_3}$, we have the following positivity constraints:
\begin{equation}
    \langle \O_{T1,1}\O_{T1,1}^\dagger \rangle \geq 0 \quad \text{and} \quad \langle \O_{T1,2}\O_{T1,2}^\dagger \rangle \geq 0,
\end{equation}
as well as any convex combinations of the above two inequalities. It is not hard to see that 
\begin{equation}
    \begin{split}
        \O_{T1,1}\O_{T1,1}^\dagger
        &= \sum_{\substack{\bk_1\bk_2\bk_3\\\bk'_1\bk'_2\bk'_3}}A_{\bk_1\bk_2\bk_3}A^*_{\bk'_1\bk'_2\bk'_3}c_{\bk_1}c_{\bk_2}c_{\bk_3}c^\dagger_{\bk'_3}c^\dagger_{\bk'_2}c^\dagger_{\bk'_1} \\
        &= \sum_{\substack{\bk_1\bk_2\bk_3\\\bk'_1\bk'_2\bk'_3}}A_{\bk_1\bk_2\bk_3}A^*_{\bk'_1\bk'_2\bk'_3}\left[ -c^\dagger_{\bk'_1}c^\dagger_{\bk'_2}c^\dagger_{\bk'_3}c_{\bk_3}c_{\bk_2}c_{\bk_1} + c^\dagger_{\bk'_3}c^\dagger_{\bk'_2}\delta_{\bk'_1,[\bk_1}c_{\bk_2}c_{\bk_3]}\right. \\
        &\qquad\qquad\qquad\qquad\qquad\quad- \left. c^\dagger_{\bk'_3}\delta_{\bk'_2,[\bk_1}c_{\bk_2}c_{\bk_3]}c^\dagger_{\bk'_1}+\delta_{\bk'_3,[\bk_1}c_{\bk_2}c_{\bk_3]}c^\dagger_{\bk'_2}c^\dagger_{\bk'_1}
        \right],
    \end{split}
\end{equation}
where $\delta_{\bk'_i,[\bk_1}c_{\bk_2}c_{\bk_3]} $ is defined as the sum of the cyclic permutations of the indices enclosed in the square bracket:
\begin{equation}
    \delta_{\bk'_i,[\bk_1}c_{\bk_2}c_{\bk_3]}\equiv\delta_{\bk'_i,\bk_1}c_{\bk_2}c_{\bk_3}+ \delta_{\bk'_i,\bk_2}c_{\bk_3}c_{\bk_1}+\delta_{\bk'_i,\bk_3}c_{\bk_1}c_{\bk_2}.
\end{equation}
Thus, the convex combination $\O_{T1,1}\O_{T1,1}^\dagger  +  \O_{T1,2}\O_{T1,2}^\dagger $ cancels exactly the three-particle terms. Then, the following constraint containing only at most two-particle terms
\begin{equation}
    \langle \O_{T1,1}\O_{T1,1}^\dagger \rangle + \langle \O_{T1,2}\O_{T1,2}^\dagger \rangle \geq 0, \quad\forall A_{\bk_1\bk_2\bk_3}
\end{equation}
is the so-called (2,3)-T1 constraint.

By using the inverse form factor matrix, we can rewrite the (2,3)-T1 constraint in terms of the two-particle structure factor:
\begin{equation}\label{2_3_T1}
    \begin{split}
        &\langle \O_{T1,1}\O_{T1,1}^\dagger \rangle + \langle \O_{T1,2}\O_{T1,2}^\dagger \rangle \\
        =& \sum_{\substack{\bk_1\bk_2\bk_3\\\bk'_1\bk'_2\bk'_3}}A_{\bk_1\bk_2\bk_3}A^*_{\bk'_1\bk'_2\bk'_3}\left[ \left( -\sum_{\bq}\Lambda^\bq_{\bk'_3,\bk_2}\Lambda^{-\bq}_{\bk'_2,\bk_3}S_\bq\delta_{\bk'_1,\bk_1} + \nu\delta_{\bk'_3,\bk_3} \delta_{\bk'_2,\bk_2}\delta_{\bk'_1,\bk_1}+ 
        \left(\text{cyclic permutation of }\bk_1,\bk_2,\bk_3\right)
        \right)\right.\\
        &- \left. \left( -\sum_{\bq}\Lambda^\bq_{\bk'_3,\bk_2}\Lambda^{-\bq}_{\bk'_1,\bk_3}S_\bq\delta_{\bk'_2,\bk_1} + \nu\delta_{\bk'_3,\bk_2} \delta_{\bk'_1,\bk_3}\delta_{\bk'_2,\bk_1}+ 
        \left(\text{cyclic permutation of }\bk_1,\bk_2,\bk_3\right)
        \right)\right.\\
        &+ \left. \left( -\sum_{\bq}\Lambda^\bq_{\bk'_2,\bk_2}\Lambda^{-\bq}_{\bk'_1,\bk_3}S_\bq\delta_{\bk'_3,\bk_1} + (2\nu-1)\delta_{\bk'_2,\bk_2} \delta_{\bk'_1,\bk_3}\delta_{\bk'_3,\bk_1}+ (1-\nu)\delta_{\bk'_2,\bk_3} \delta_{\bk'_1,\bk_2}\delta_{\bk'_3,\bk_1}+
        \left(\text{cyclic permutation of }\bk_1,\bk_2,\bk_3\right)
        \right)
        \right]\\
        \geq & 0, \quad\forall A_{\bk_1\bk_2\bk_3}\in\mathbb{C}.
    \end{split}
\end{equation}
Like what has been discussed in the (2,2) positivity case, we can also antisymmetrize the indices to reduce the dimensionality of the matrix.

Similar to the (2,3)-T1 constraint discussed above, there is another simple (2,3)-positivity constraint bared the name (2,3)-T2 constraint as an extension to the T1, which involves the following two operators: 
\begin{equation}
    \begin{split}
        \O_{T2,1} =& \sum_{\bk_1,\bk_2,\bk_3}A_{\bk_1\bk_2\bk_3}c_{\bk_1}c_{\bk_2}c^\dagger_{\bk_3}+\sum_\bk B_\bk c_{\bk} \\
        \O_{T2,2} =& \sum_{\bk_1,\bk_2,\bk_3}A^*_{\bk_1\bk_2\bk_3}c^\dagger_{\bk_1}c^\dagger_{\bk_2}c_{\bk_3}+\sum_\bk D^*_\bk c^\dagger_{\bk},
    \end{split}
\end{equation}
which are really the generalized T2 operators.
Following the same reasoning discussed above, we have $\O_{T2,1}\O_{T2,1}^\dagger  +  \O_{T2,2}\O_{T2,2}^\dagger $ cancels exactly the three-particle terms. Writing the (2,3)-T2 constraint using the structure factor and the inverse form factors is similar to what we did for the T1 case but more complicated, which will be omitted here. This procedure can then be extended to any (2,p)-positivity constraints.

\end{document}